\documentclass[referee]{raa}            

\usepackage{graphicx,times}             
\usepackage{natbib}
\usepackage{amssymb,amsmath}
\bibpunct{(}{)}{;}{a}{}{,}
\usepackage{multirow}

\usepackage[pagebackref=true]{hyperref}
\begin{document}

  \title{Oxygen-rich Mira variables in the Large Magellanic Cloud}

   \volnopage{Vol.0 (202x) No.0, 000--000}      
   \setcounter{page}{1}          

   \author{Do T. Hoai
      \inst{*,1,2}\footnotetext{$^*$Corresponding Authors, these authors contributed equally to this work.}
   \and Pham. T. Nhung
      \inst{*,1}
      \and Pierre Darriulat
      \inst{*,1}
   }

   \institute{$^1$ Vietnam National Space Center (VNSC), Vietnam Academy of Science and Technology (VAST), 18 Hoang Quoc Viet, Nghia Do, Ha Noi, Vietnam.\\
   $^2$ University of Science and Technology of Hanoi (USTH), Vietnam Academy of Science and Technology (VAST), 18 Hoang Quoc Viet, Nghia Do, Ha Noi, Vietnam.\\
\vs\no
   {\small Received 202x month day; accepted 202x month day}}

   \abstract{A study of the shapes of the light curves of oxygen-rich Mira variables in the Large Magellanic Cloud taken from the OGLE data-base is presented. The comparison with curves observed in the solar neighbourhood in the 250- to 450-day-period interval confirms the presence of two different families, displaying similar properties in their evolution along the AGB. Most of these stars are shown to become carbon-rich before having their period reach 450 days. Higher period curves are identified as being emitted by Hot Bottom Burners and OH/IR stars.
     }

   \authorrunning{D.T. Hoai, P.T. Nhung \& P. Darriulat}            
   \titlerunning{}  

   \maketitle

%
%
\section{Introduction}
In a recent study \citep{Darriulat2026, Hoai2026} of the relation between visible band light curves and evolution along the AGB in the solar neighbourhood, we revealed differences between two families of stars, which we called A and B, most likely associated with larger ($\sim$2 to 4 M$_{\odot}$) and smaller ($\sim$0.5 to 2 M$_{\odot}$) initial masses, respectively. The availability of accurate data from OGLE \citep{Soszynski2009a} offers an opportunity to look at this question in the Large Magellanic Cloud (LMC), which we do in the present article.

\begin{figure*}
  \centering
  \includegraphics[width=0.7\textwidth,trim={0cm 0cm 0cm 0},clip]{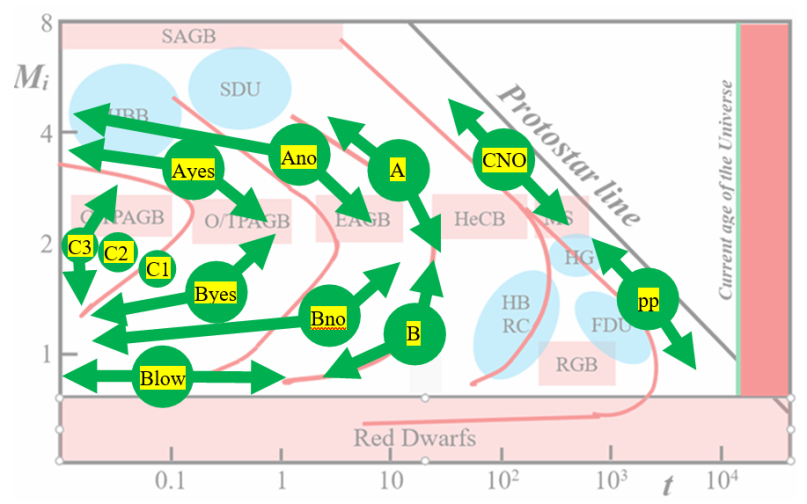}
  \caption{Schematic evolution of AGB stars in the initial mass ($M_i$, measured in M$_{\odot}$) vs time ($t$, measured backward from death in Myr) as described in \citet{Darriulat2026}.}
  \label{fig1}
\end{figure*}

In summary, the earlier work considered separately the progenies of stars that enter the Asymptotic Giant Branch (AGB) already pulsating with amplitudes at the scale of one unit of magnitude, referred to as the A family, and the progeny of stars that enter the AGB with little variability, with amplitudes not exceeding $\sim$0.4 unit of magnitude, referred to as the B family. Both have early M spectral types and temperatures above 3500 K. Their evolution is schematically displayed in Figure \ref{fig1}.

The A family, leaving aside stars with initial masses larger than some 4 M$_{\odot}$ which become Hot Bottom Burners (HBB), includes stars that spend their whole AGB lives as Mira variables, climbing the Early AGB (EAGB) with their light curves deviating progressively from a sine wave, their temperatures decreasing from above 3500 K to some 3000 K and their periods increasing to some 200 days. They then keep expanding and cooling with increasing periods and decreasing widths of the light pulse. Some may reach temperatures below 2500 K and periods over 500 days without entering the TPAGB, or at least, if they enter it, without experiencing strong Third Dredge Up (TDU) events; these keep having a modest mass-loss rate at the scale of 10$^{-7}$ M$_{\odot}$\,yr$^{-1}$ and their light curves display narrower and narrower light pulses; they may then leave the AGB as Tc-poor oxygen-rich stars having a low isotopic $^{12}$C/$^{13}$C ratio, below $\sim$20. But many will enter the Thermally Pulsing AGB (TPAGB) at various stages of their expansion and cooling-down phase, becoming Tc-rich stars of the groups named My1 and S1 in \citep{Hoai2026} with a carbon isotopic ratio in excess of 20, and reaching temperatures at the 2000 K level or below and mass-loss rates well in excess of 10$^{-6}$ M$_{\odot}$\,yr$^{-1}$, possibly becoming cool carbon-rich stars before leaving the AGB.

In contrast, the stars of the B family maintain a low variability when evolving along the EAGB. They do not start displaying significant pulsations before reaching spectral type M4. Then, some, having spectral types M4 to M6 will start pulsating with amplitudes soon reaching 2 units of magnitude and typical periods of 200 to 300 days, before entering the TPAGB; they still have low mass-loss rates, at the scale of 10$^{-7}$ M$_{\odot}$\,yr$^{-1}$. The others, which are likely to have low initial masses, below some 1.5 M$_{\odot}$, will keep expanding and cooling down without experiencing strong TDU events, ultimately leaving the AGB as oxygen-rich stars with low amplitudes not exceeding 0.6 mag. The former will then keep pulsating with increasing amplitudes and periods, eventually entering the TPAGB and becoming Mira variables experiencing strong TDU events, corresponding to groups My2, S2 and S3 of Hoai et al. (2026). On the TPAGB, both the Mira variables of the A and B families (Figure \ref{fig1}) have spectral types close to M6 and display features that give evidence for their having experienced strong TDU events: they are Tc-rich and their $^{12}$C/$^{13}$C isotopic ratio has reached $\sim$30; but those of the B family differ clearly from those of the A family as having a mean temperature $\sim$400 K higher, a mean period $\sim$100 days shorter, a significantly lower mass-loss rate and remarkably broader light pulses. As a result, those which become carbon-rich enter different families: the carbon-rich progeny of the A family have temperatures below $\sim$2400 K while that of the B family have temperatures above it. 
  
The detailed inspection of the light curves of a large number of semi-regular variables, as has been done in \citet{Darriulat2026}, gives evidence for a very broad variety of irregularities; amplitudes may change progressively over many cycles or abruptly over a few; the same applies for periods, however less frequently. In most cases such changes occur at times that succeed each other irregularly. The exercise gives the strong impression of a very complex and disorganized picture, leaving little hope for describing such light curves in simple terms. In particular, their study cannot be reduced to identifying the dominant mode and sequence in a period vs luminosity diagram; for such a description to tell us about the physics that governs the pulsation, one would need to identify each mode with a term of the development of the physics equation of relevance, but we are far from being able to do so. Indeed, the current understanding of the dynamical processes at play \citep[][and references therein]{Ahmad2023, Ahmad2025} shows the importance of the non-linearity of the interplay between the self-excited pulsations and the self-consistent convection and the natural emergence of multi-mode pulsations, including both radial and non-radial modes. In such a situation it is therefore essential to keep one's mind open to other possible descriptions and to pay attention to what may look as exotic details to someone who is not familiar with the scrutiny of such light curves.

The OGLE data are of a different nature from the AAVSO data used in our earlier work. A much shorter time span (one decade instead of one century), I-band (near infrared) rather than visible, presence of gaps in the light curves; moreover, the stars in the Magellanic clouds have a different history than those in the Galaxy, in particular with a metallicity about half of solar for the LMC, implying a longer stay as carbon-rich (Table \ref{tab1}). Their study requires therefore a different approach than used in our earlier work. Another major difference is that all LMC stars are at a same known distance from us while obtaining reliable measures of the distances of stars in the solar neighbourhood is often difficult.

\begin{table*}
  \centering
  \caption{Effect of the lower metallicity for LMC than for Galaxy stars \citep[from][]{Miller2016}. The left column lists the initial mass in units of solar masses. The other columns compare the durations of the presence in the Main Sequence (MS), Red Giant Branch (RGB), He Core Burning phase (HeCB), early Asmptotic Giant Branch (eAGB), Thermally Pulsing Asymptotic Giant Branch (TPAGB), either oxygen-rich (spectral type M) or carbon rich (spectral type C). The two numbers given as $x$/$y$ in the table correspond to metallicities 0.02 ($x$) and 0.01 ($y$), respectively. Durations differing by more than 50\% are shown in bold.}
  \label{tab1}
  \begin{tabular}{ccc ccc c}
    \hline
    $M_i$& MS &RGB &HeCB & eAGB  & TPAGB/M  &TPAGB/C  \\
   (M$_{\odot}$)& (10$^9$ yr) &(10$^7$ yr) &(10$^7$ yr) & (10$^6$ yr) & (10$^5$ yr) &(10$^4$ yr) \\
    \hline
1 & 9.6/7.9
&204/170
&13/11
&11/12
&6.6/7.4
&0/0\\
2
&1.1/0.9
&3.4/2.7
&31/28
&19/13
&\bf{29/16}
&\bf{10/38} \\
3
&0.3/0.3
&0.47/0.41
&8.4/7.3
&\bf{6.2/4.1}& \bf{8/3}
&11/14\\
\hline
\end{tabular}
\end{table*}

Earlier work on the OGLE LMC observation of Long Period Variable stars (LPV) has focussed on studies of the period-luminosity relations that they obey \citep{Soszynski2005, Soszynski2009a, Ita2011}, often in conjunction with infrared observations from other sources, allowing for colour evaluations. Such is the case of the work of \citet{Hey2025} using ATLAS observations, of \citet{Groenewegen2004} using 2MASS and Denis observations, of \citet{Feast1989} using observations of the 1.9 m SAAO telescope, of \citet{Fraser2005} using MACHO and 2MASS observations, of \citet{Ita2002, Ita2004a, Ita2004b} using Sirius observations, of \citet{Yuan2017} using Near Infrared Synoptic Survey (LMCNISS) observations. Other studies of the LPVs in the LMC include the work of \citet{Spano2011}, who report on EROS-2 survey observations and of \citet{Riebel2015} and \citet{Prager2022} on observations of the Surveying the Agents of Galaxy Evolution (SAGE)-Var program. \citet{Wen2024} use observations from UKIRT/WFCAM and 2MASS to study the mass-loss rates. Most of these studies pay particular attention to the properties of the circumstellar dust and the obscuration that results, as mentioned early by \citet{Wood1992, Wood2000} and \citet{Whitelock2003}. \citet{Lebzelter2019} and \citet{Trabucchi2019} use Gaia and 2MASS observations to perform a very detailed study of the evolution of the LMC LPVs in the space spanned by different infrared bands and reveal the existence of an offset between low- and intermediate-mass oxygen-rich stars for both fundamental and first overtone in agreement with simulations. 

\section{Mira variables}
\subsection{Selecting clear pulsations}
The distribution of the OGLE data between Miras (defined as having a mean amplitude of oscillation larger than 0.8 mag) and SRVs (smaller than 0.8 mag) and between carbon-rich and oxygen-rich (distinguished from their different colour properties) is shown in Table \ref{tab2} as a function of period. The precise definition of such grouping of curves is given by \citet{Soszynski2009a} and is meant to help their study but should not be understood as defining rigorously different physical categories. For oxygen-rich stars, Miras dominate over SRVs when the period exceeds ~300 days; and for carbon-rich stars, when it exceeds $\sim$400 days.

In order to study the light curves of the stars of the OGLE data-base, we need to describe them in terms of parameterisations that help with understanding the underlying mechanism of pulsation. This requires selecting curves that display clear pulsations.The parameterisation offered by the OGLE catalogue consists of an evaluation of the mean I-band luminosity, $I$, and of the six parameters, periods and amplitudes, defining the three dominant modes obtained from a Fourier analysis, $P_1$/$P_2$/$P_3$ and $A_1$/$A_2$/$A_3$. Particular care has been given to the evaluation of the dominant mode parameters, $P_1$ and $A_1$. When the first mode dominates clearly over the other two, we can expect them to give reliable measures of the mean period and amplitude and the shape of the light curve to be reasonably well defined. Visual inspection of a very large number of curves has shown that this is indeed often the case, but not always.

\begin{figure*}
  \centering
  \includegraphics[width=0.3\textwidth,trim={0cm 0cm 0cm 0},clip]{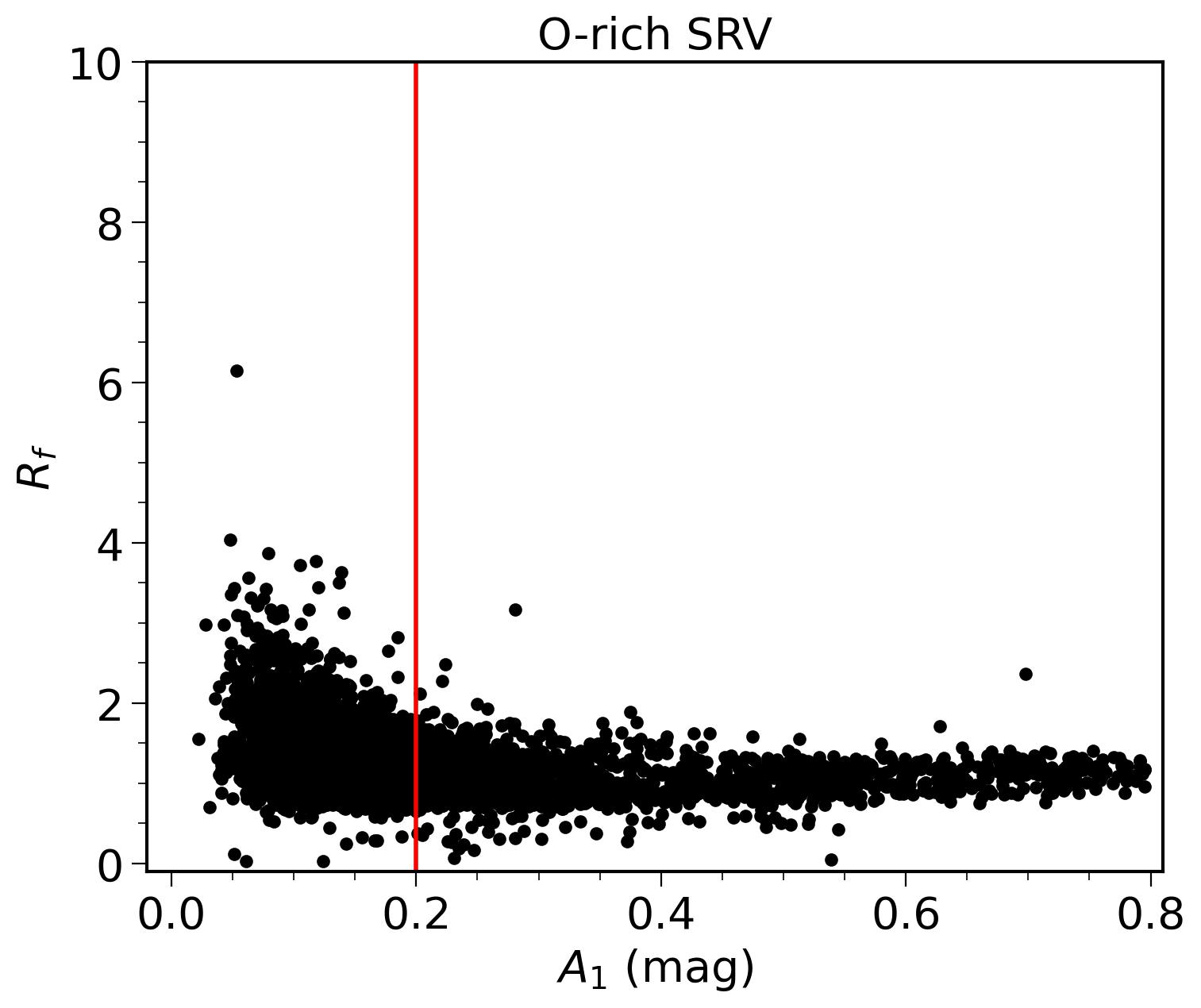}
  \includegraphics[width=0.3\textwidth,trim={0cm 0cm 0cm 0},clip]{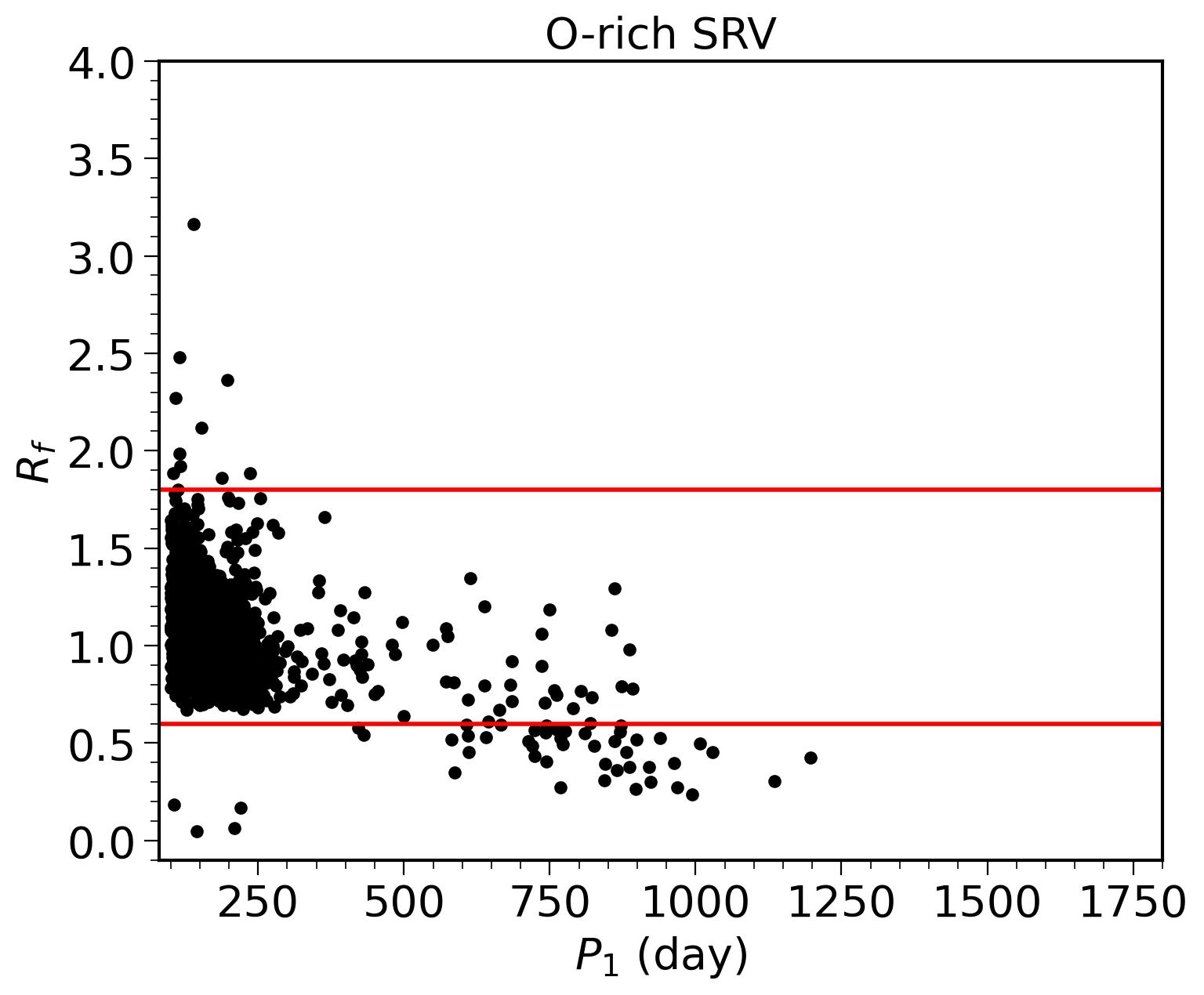}
  \includegraphics[width=0.3\textwidth,trim={0cm 0cm 0cm 0},clip]{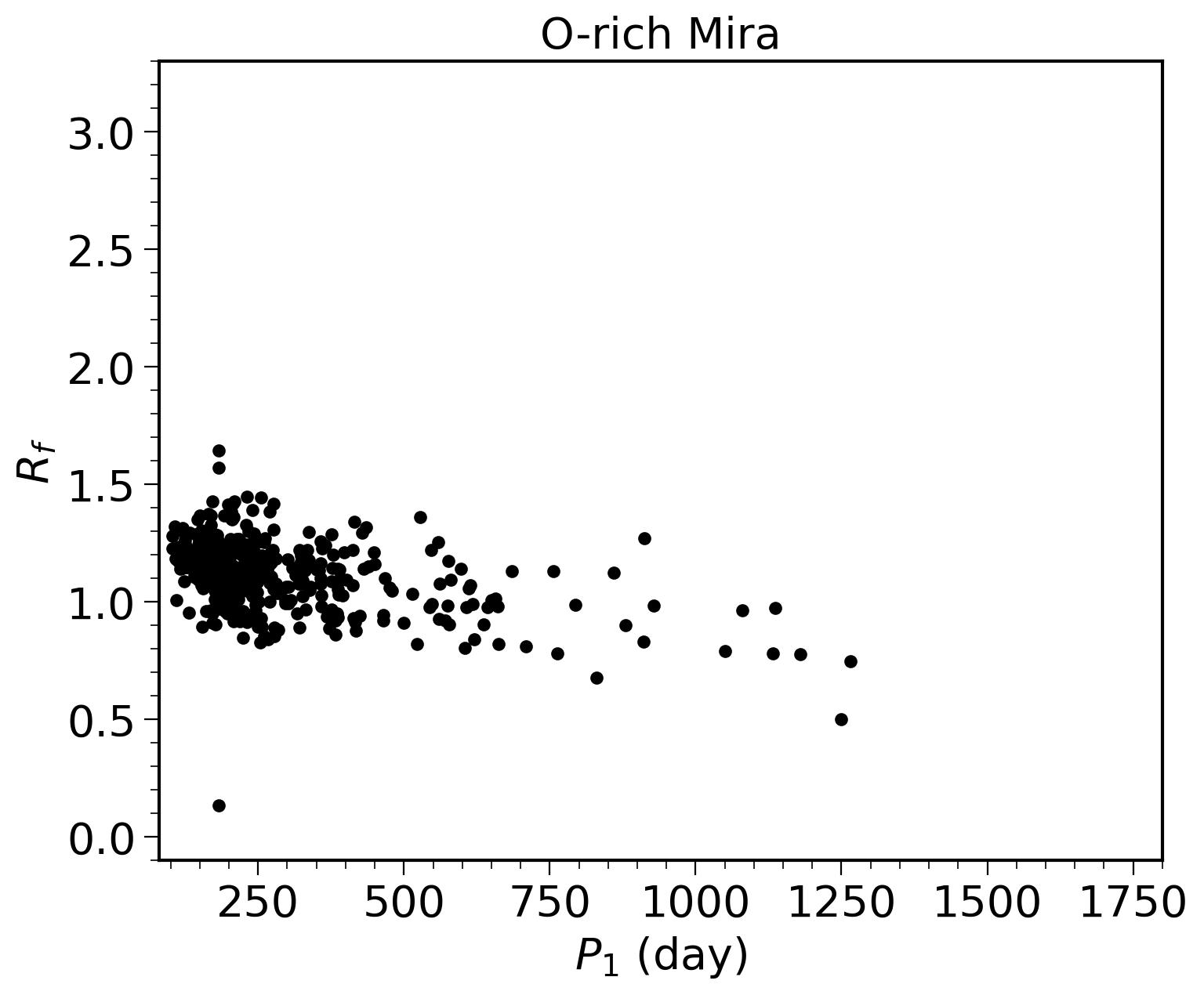}
  \includegraphics[width=0.3\textwidth,trim={0cm 0cm 0cm 0},clip]{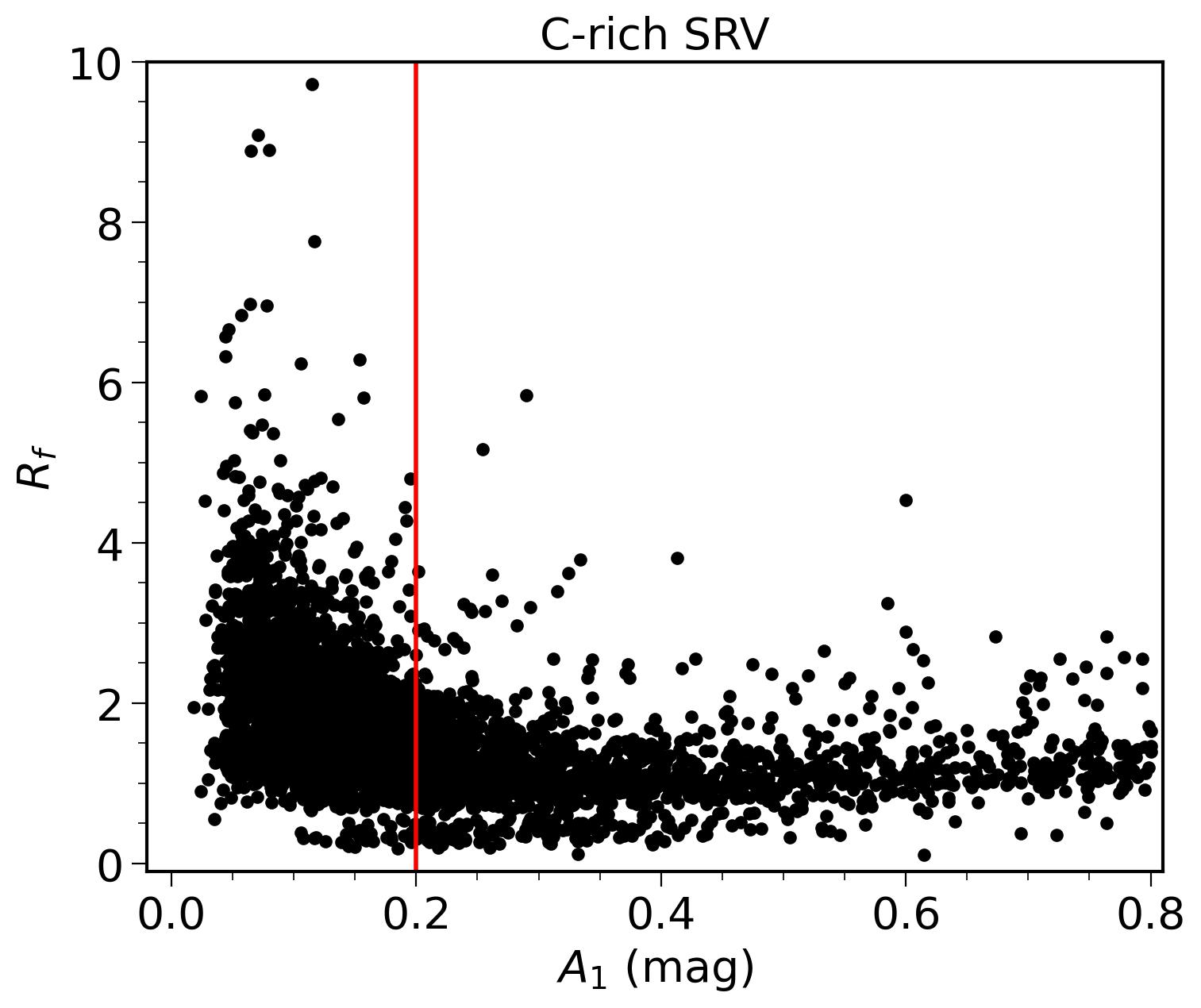}
  \includegraphics[width=0.3\textwidth,trim={0cm 0cm 0cm 0},clip]{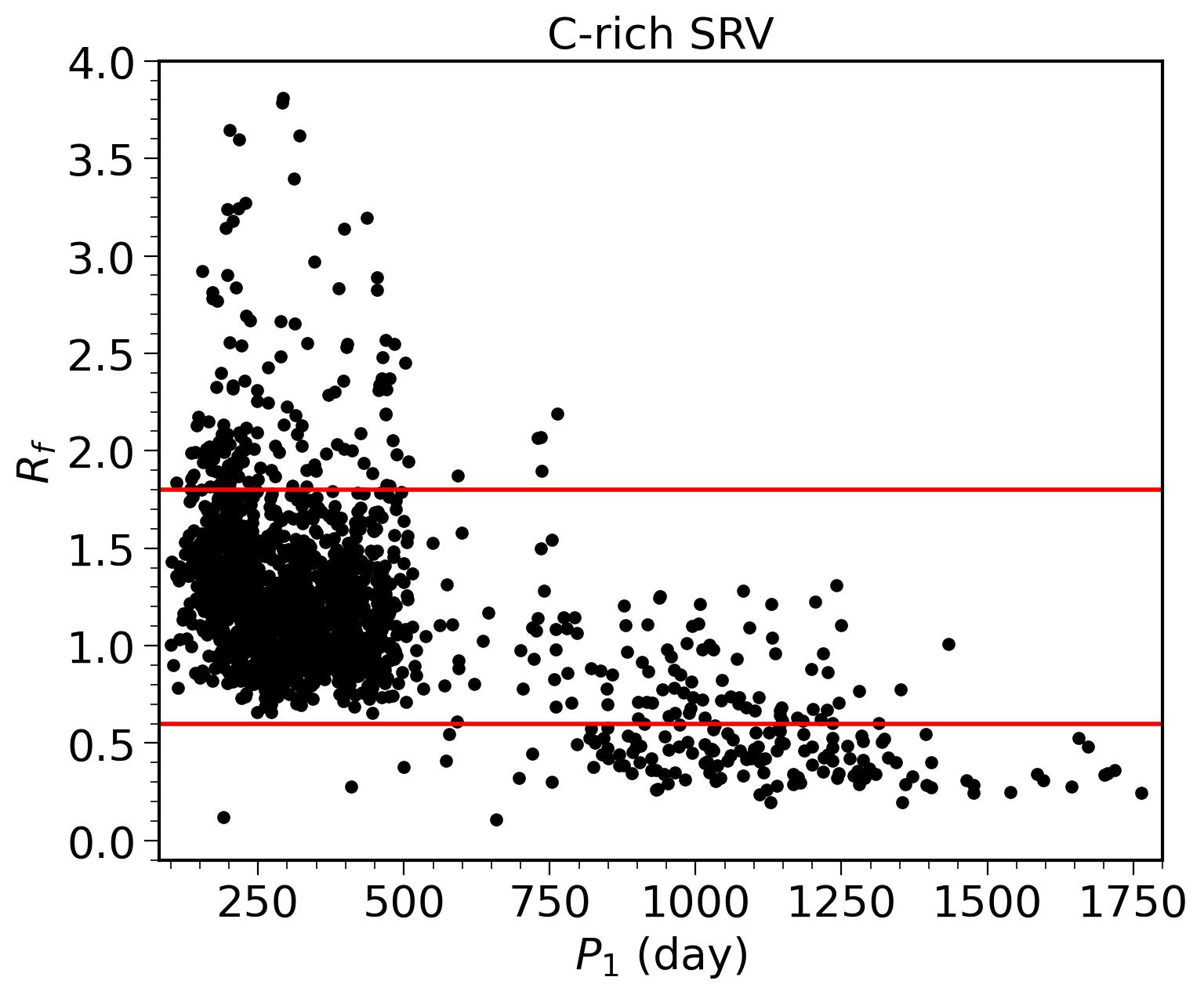}
  \includegraphics[width=0.3\textwidth,trim={0cm 0cm 0cm 0},clip]{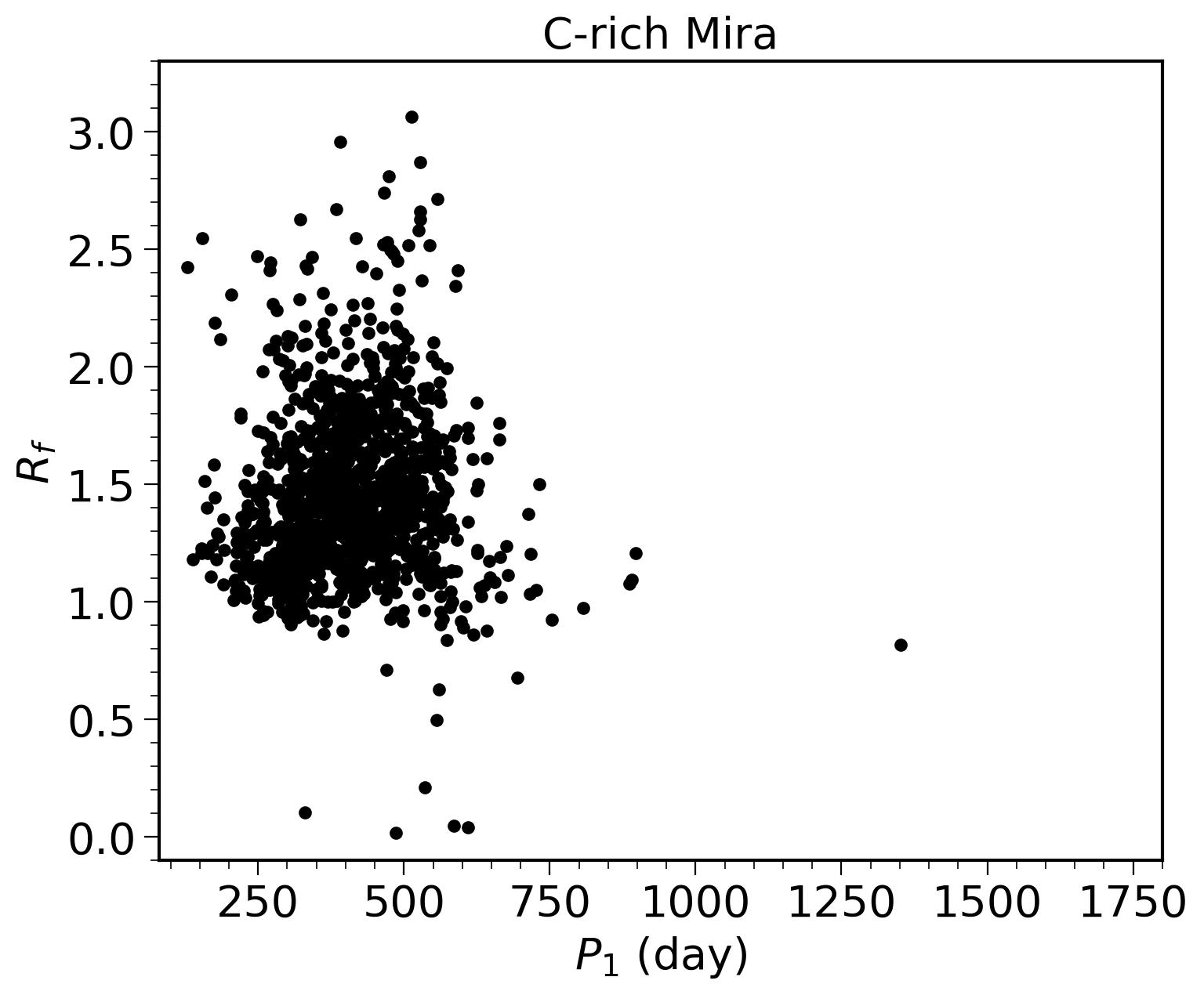}
  \caption{Properties of the parameter $R_f$ for oxygen-rich (upper row) and carbon-rich (lower row) curves of the LMC/AGB OGLE data-base having $P_1$$>$100 days. The left panels show the dependence of $R_f$ on $A_1$. The red line indicates the cut applied to all curves in the present study. The central and right panels show, after application of the cut, the dependence of $R_f$ on $P_1$ for SRV and Miras, respectively.}
  \label{fig2}
\end{figure*}

In order to decide to which extent $P_1$ and $A_1$ can, to a first approximation, provide a sensible description of the properties of the light curve, we define a parameter, $R_f$, meant to be a good proxy of such ability. It is defined as the ratio between the frequency of the dominant mode, $f_1=1/P_1$ and the mean frequency $f_0=(A_1/P_1+A_2/P_2+A_3/P_3)/(A_1+A_2+A_3)$. If $A_2$ and $A_3$ are much smaller than $A_1$, or if $P_2$ and $P_3$ are close to $P_1$, $R_f$ is close to 1. Figure \ref{fig2} displays its dependence on $A_1$ for the oxygen-rich and carbon-rich curves of the OGLE data-base having $P_1$$>$100 days. Together with a detailed visual inspection of some of the curves, it shows that requiring $A_1$$>$0.2 mag selects curves for which an evaluation of the shape can be done reliably. The curves rejected by the cut are too irregular to be reliably described in terms of just a few parameters. In general, they are not chaotic, but display irregularities such that reducing their description to the definition of periods and amplitudes of dominant modes cannot be expected to be of much help. We therefore apply this cut to all the data used in the present study. The resulting dependence of $R_f$  on $P_1$ is shown in Figure \ref{fig2} for Miras and SRV and for oxygen-rich and carbon-rich curves separately. The difference observed between oxygen-rich and carbon-rich is spectacular, implying the need for a detailed understanding of its meaning. As the present article is limited to the study of the oxygen-rich sample, we defer this discussion to a later presentation of the study of the carbon-rich sample (Nhung et al., in preparation). For what concerns the oxygen-rich sample, the curves, particularly for the Miras, are generally well grouped together close to $R_f$$=$1. 

\begin{table*}
  \centering
  \caption{Distribution of the LMC AGB stars with period $P_1$$>$100 days}
  \label{tab2}
  \begin{tabular}{ccc cc}
    \hline
    \multirow{2}{*}{$P_1$ (days)}  & \multicolumn{2}{c}{Carbon-rich}& \multicolumn{2}{c}{Oxygen-rich} \\
    \cline{2-5}
&Mira&SRV&Mira&SRV\\
\hline
100-150
&2
&1144
&56
&1966 \\
150-200
&18
&1315
&126
&643\\
200-250
&51
&648
&120
&380 \\
250-300
&117
&447
&47
&67 \\
300-400
&439
&592
&55
&29 \\
400-500
&354
&283
&20
&22 \\
500-600
&177
&32
&15
&15 \\
600-700
&28
&7
&13
&16 \\
700-800
&6
&27
&4
&21 \\
800-900
&4
&32
&3
&20 \\
900-1000
&0
&56
&3
&6 \\
$>$1000
&1
&150
&7
&5 \\
\hline
Total
&1197
&4733
&469
&3190 \\
\hline
\end{tabular}
\end{table*}

\begin{figure*}
  \centering
  \includegraphics[width=0.9\textwidth,trim={0cm 0cm 0cm 0},clip]{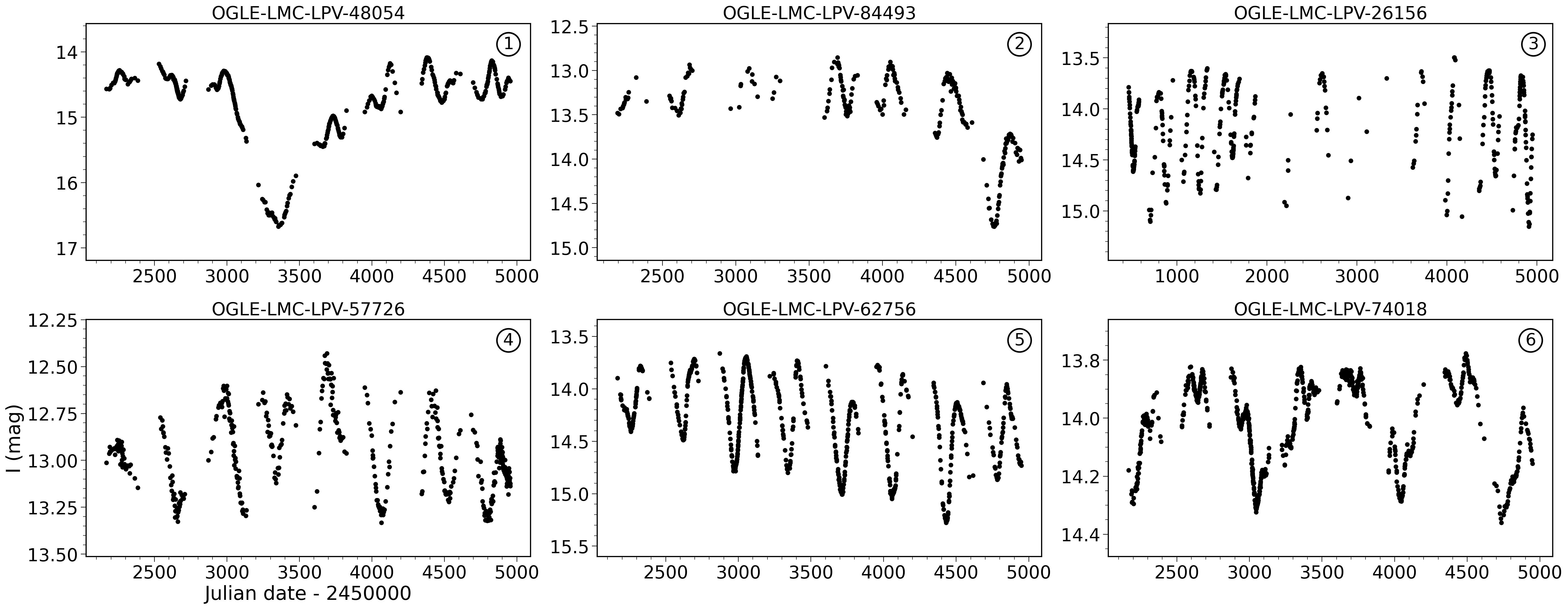}
  \caption{Examples of oxygen-rich outliers of the $R_f$ distribution. From left to right and from up down, panels 1 and 2 show SRV curves presenting an episode of obscuration mimicked by $P_2$$=$2179 days and $P_3$$=$3333 days, respectively; panel 3 shows a Mira curve lacking observations at central time, mimicked by $P_2$$=$3759 days; panels 4 and 5 show an SRV and a Mira curves displaying variable amplitudes of oscillation, mimicked by $P_2$$=$2817 days and 3759 days, respectively. The right panel is typical of the group of SRV curves having $P_1$$>$$\sim$500 days; it has $P_1$$=$899 days, $P_2$$=$171 days, $P_3$$=$692 days, $A_1$$=$0.33 mag, $A_2$$=$0.11 mag, $A_3$$=$0.09 mag and $I$$=$14.1 mag.}
  \label{fig3}
\end{figure*}

It is instructive at this stage to look at the oxygen-rich outliers in the upper row of Figure \ref{fig2}, in order to understand what may cause $R_f$ to deviate from unity. In only a very few cases, an abnormally low density of observations prevents any meaningful analysis of the light curve to be performed. Most of the oxygen-rich SRV curves having $P_1$$>$$\sim$500 days take the form of Long Secondary Period curves, as displayed in the lower-right panel of Figure \ref{fig3}; a better description is in terms of very irregular variations at the scale of 0.1 mag, intermittently obscured by dips of light at the scale of 0.5 mag; as there are good arguments pleading for these stars being in fact carbon-rich but surrounded by only little dust, which would cause their OGLE assignment as oxygen-rich, we defer their study to a future publication reporting on the carbon-rich Long Period Variables of the LMC (Nhung et al., in preparation). In the near totality of the other cases, one of $P_2$ or $P_3$, generally $P_2$, takes a very large value, often exceeding the total observation time of 2900 days, in order to mimic some peculiarity displayed by the light curve. Figure \ref{fig3} illustrates a few typical examples. Indeed, episodes of obscuration and variations of the oscillation amplitude are features commonly met in the present study, and even more in the study of the carbon-rich sample; they require a proper and reliable interpretation. In all cases, independently from the value taken by $R_f$, the values of $P_1$ and $A_1$ quoted by OGLE match well what is observed by visual inspection: they can be used with confidence as reliably describing the main behaviour of the curves. Of course, this does not mean that they describe physical pulsations, they may rather describe intermittent obscuration.  
     
\subsection{Describing the shapes of the curves and collecting relevant information} 
The periods of Galaxy Mira variables identified in \citet{Darriulat2026} as distinguishing between families A and B, in particular between two groups referred to as My1/S1 and My2/S2/S3 in our earlier work \citep{Hoai2026}, cover approximately a 250-day-wide range, between $\sim$250 and $\sim$500 days. Figure \ref{fig2} shows that in such a range, and more generally for periods larger than 250 days, we can expect to obtain reliable descriptions of the shape of the oxygen-rich Mira light curves in the LMC. Therefore, in a first step, we restrict the study to the 167 curves selected by OGLE as ``LMC, LPV, Miras, AGB, Oxygen-rich'' having a period $P_1$$>$250 days. Visual inspection of the curves shows that 14 of these do not display data of a sufficient quality to allow for a reliable definition of their shapes: we exclude them from the sample under study and the remaining 153 retained curves are listed in Table \ref{tabA1} and illustrated in Figures \ref{figA1} to \ref{figA7} of the appendix.

\begin{figure*}
  \centering
  \includegraphics[width=0.9\textwidth,trim={0cm 0cm 0cm 0},clip]{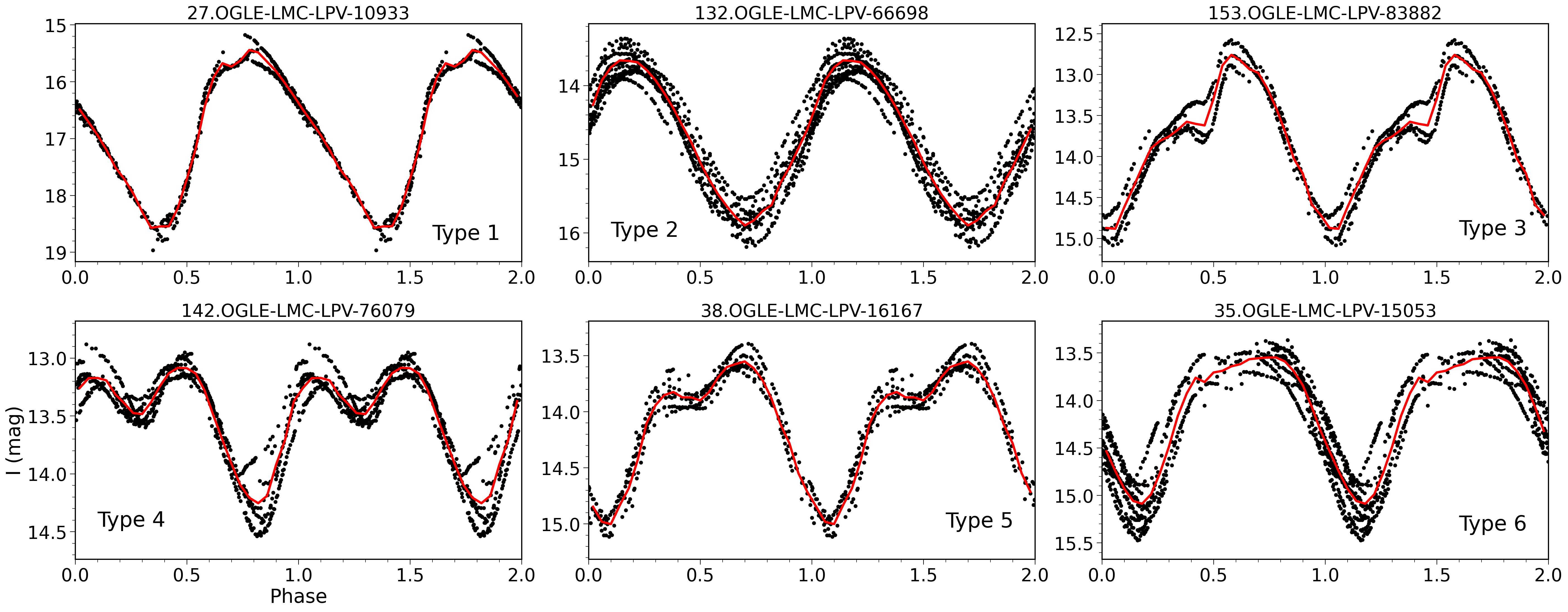}
  \caption{From left to right and from up down, typical mean profiles of types 1 to 6 as defined in the text.}
  \label{fig4}
\end{figure*}

Visual inspection of the selected light curves reveals the existence of different types of pulse profiles illustrated in Figures \ref{fig4} and $A_2$ to $A_7$; we arbitrarily define six different types of profiles as providing a reasonable description of the main features. This is nothing more than a subjective exercise only meant to help with the study; other definitions are of course possible. Type 1 includes asymmetric pulse profiles with a much steeper ascending branch than the descending branch. Type 2 includes instead symmetric type profiles close to sine waves. Type 3 includes a small group of curves displaying a clear hump at mid-height followed by a steep ascending branch. Type 4 includes curves displaying a clear double maximum. Type 5 includes the very large group of curves having a broad pulse profile with the maximum being flat on its first half and peaking on its second half. Finally, type 6 includes broad, but relatively flat, pulse profiles. In several cases the assignment of a profile type is not obvious: in such cases, as seen in Figures $A_2$ to $A_7$, a given profile might equally well be assigned to one or another of the six types. 

\begin{figure*}
  \centering
  \includegraphics[width=0.33\textwidth,trim={0cm 0cm 0cm 0},clip]{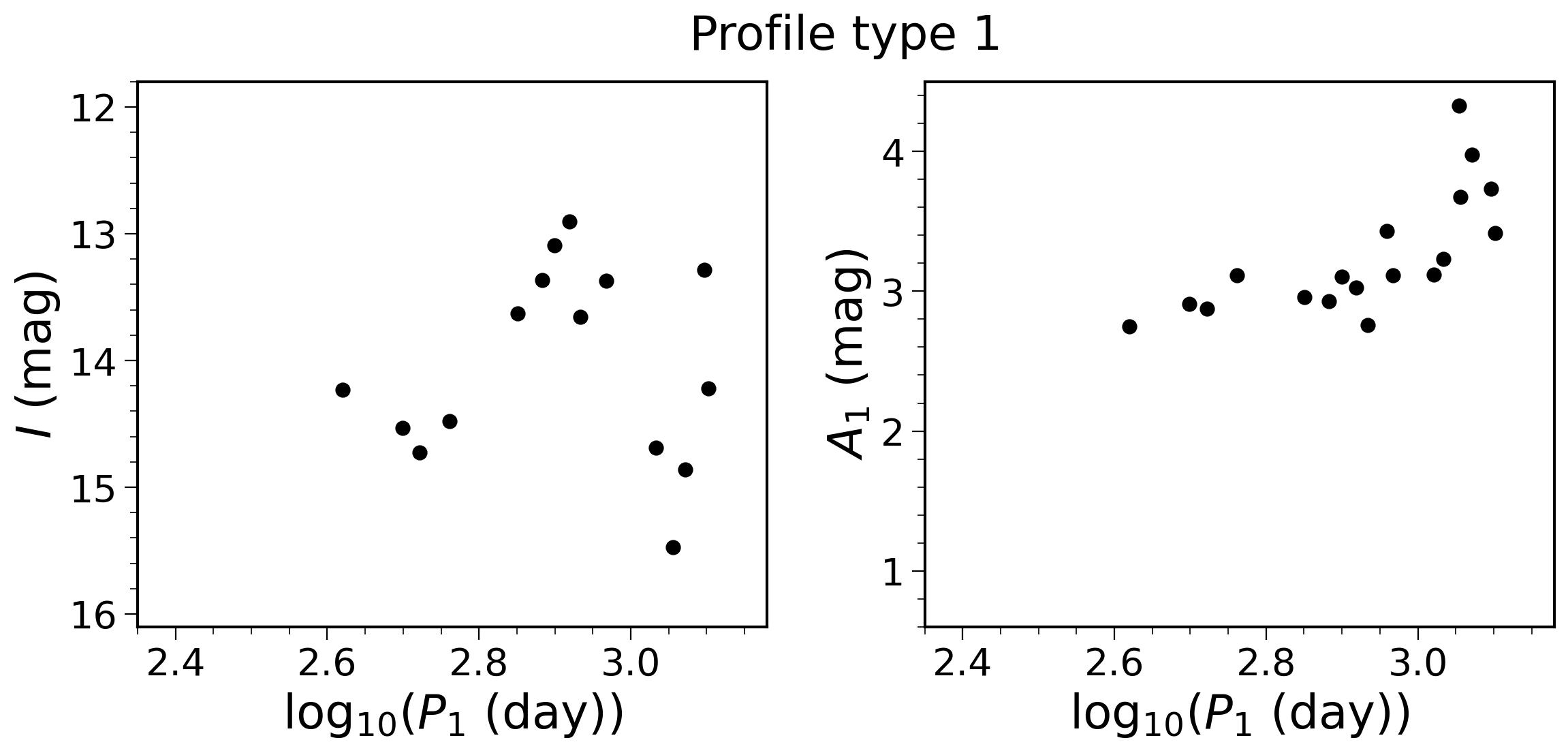}
  \includegraphics[width=0.33\textwidth,trim={0cm 0cm 0cm 0},clip]{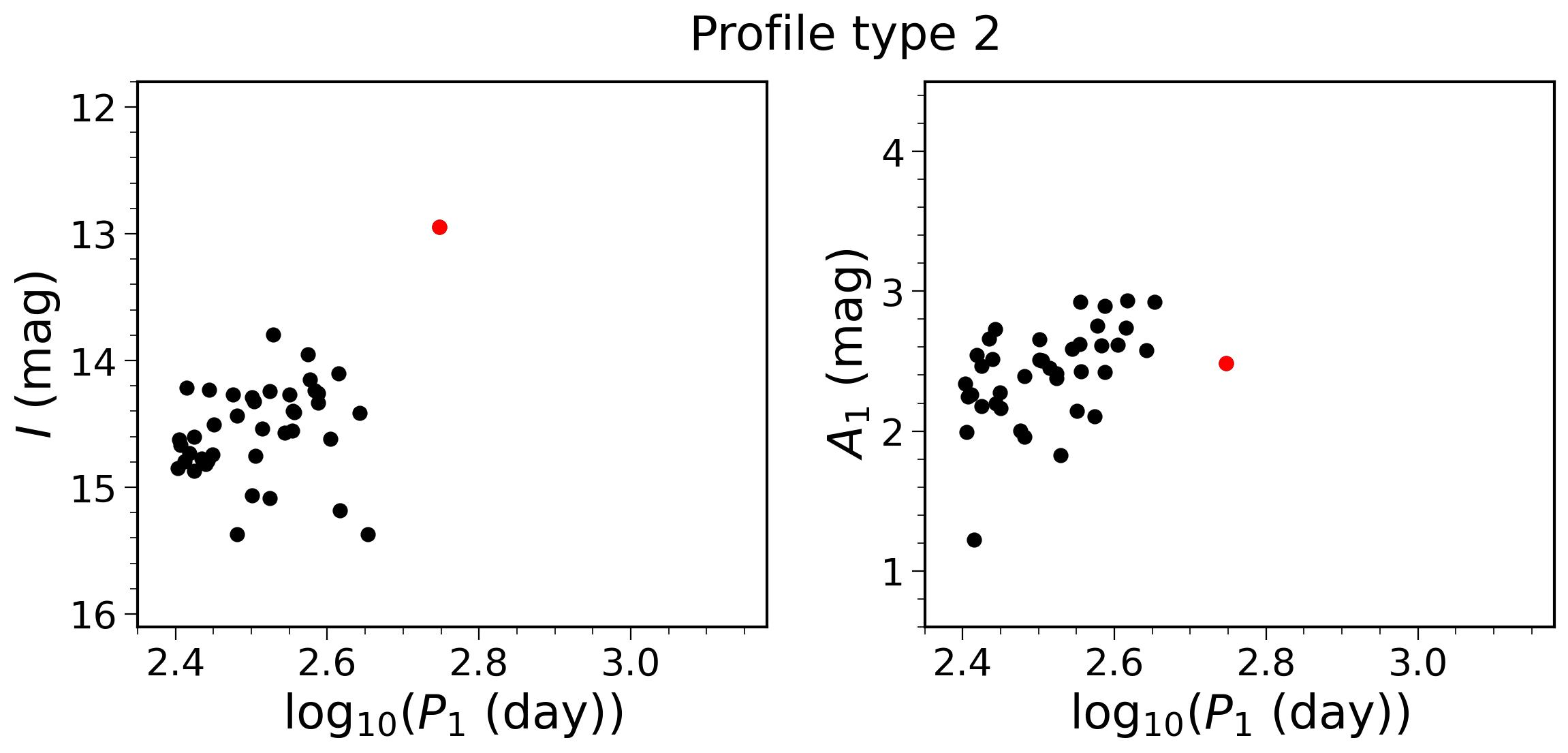}
  \includegraphics[width=0.33\textwidth,trim={0cm 0cm 0cm 0},clip]{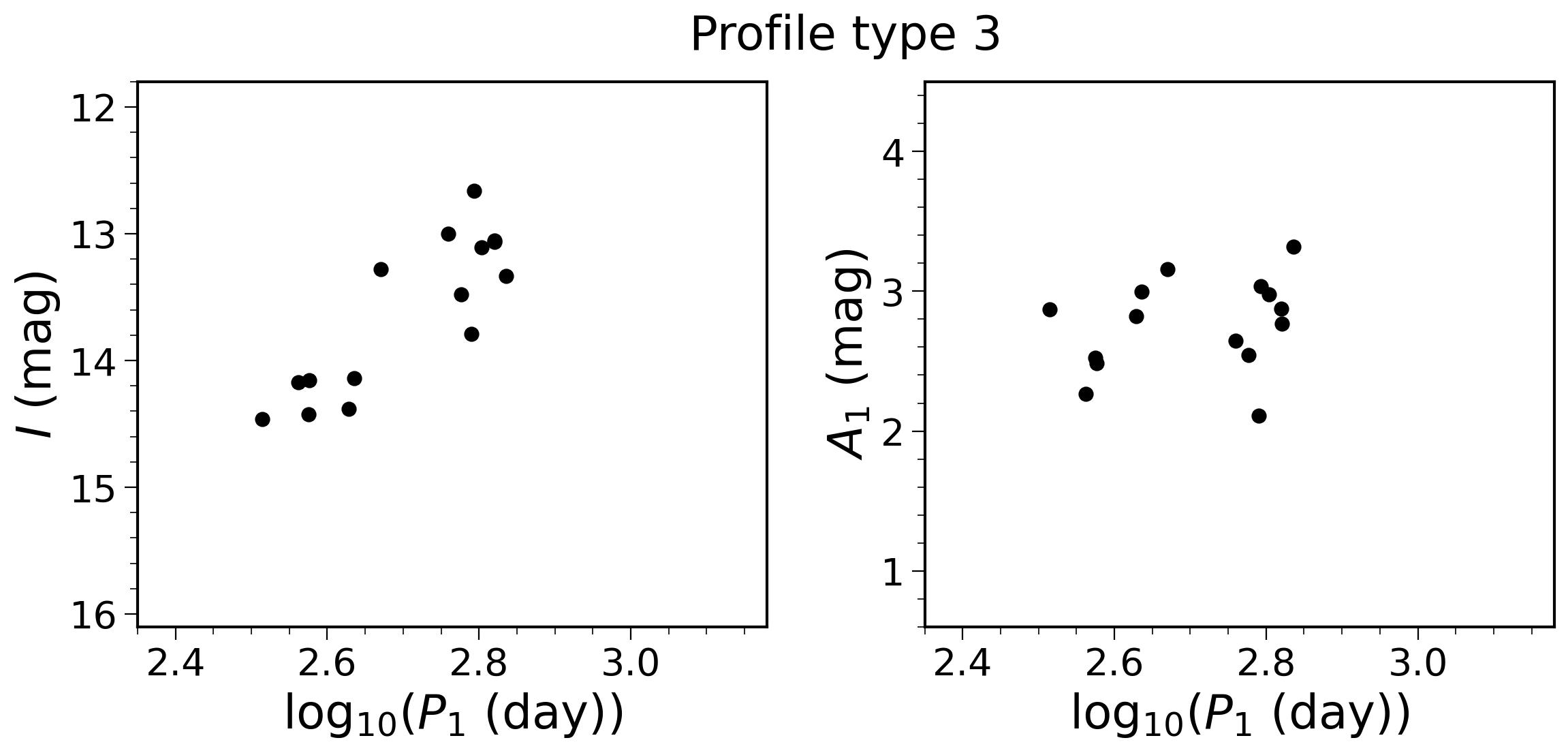}
  \includegraphics[width=0.33\textwidth,trim={0cm 0cm 0cm 0},clip]{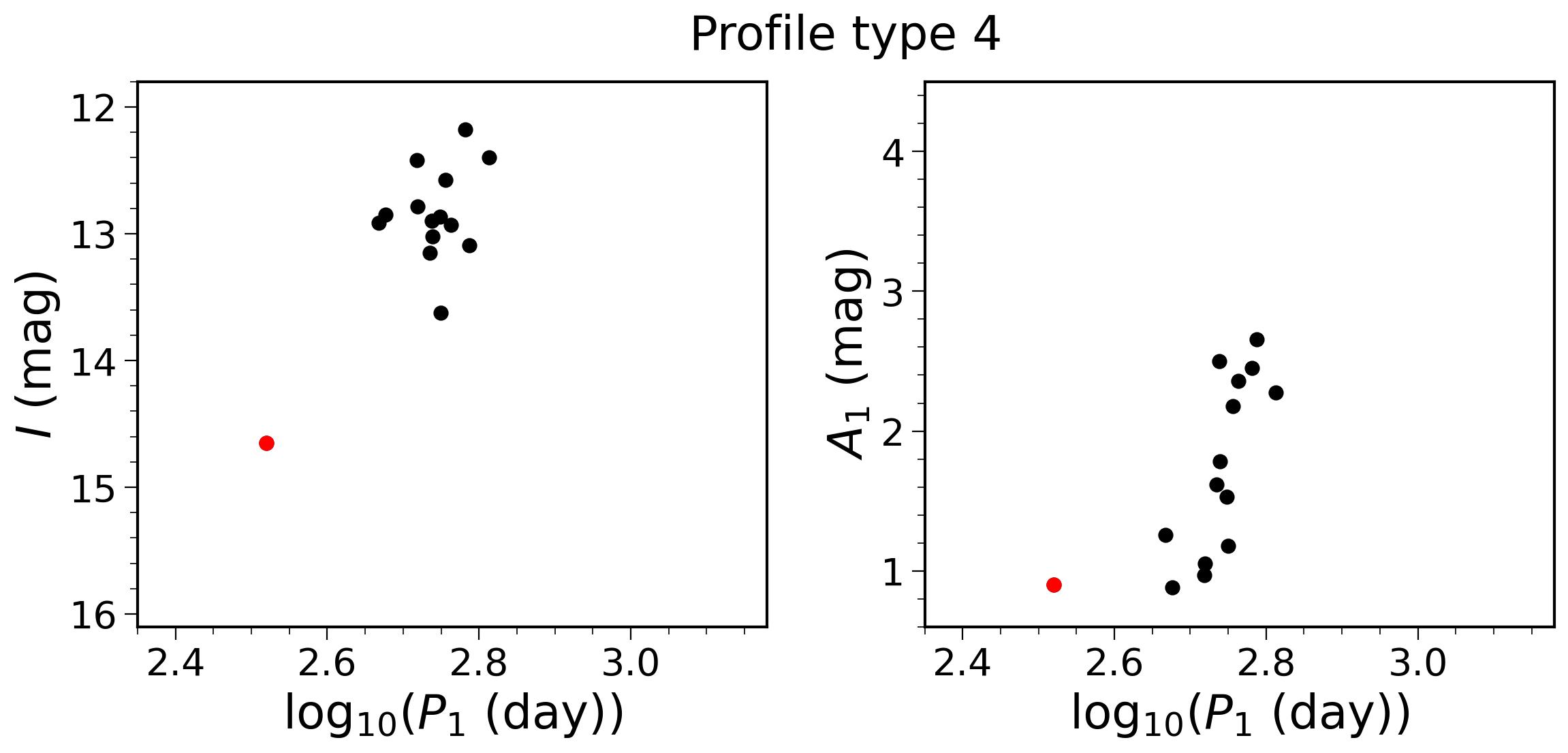}
  \includegraphics[width=0.33\textwidth,trim={0cm 0cm 0cm 0},clip]{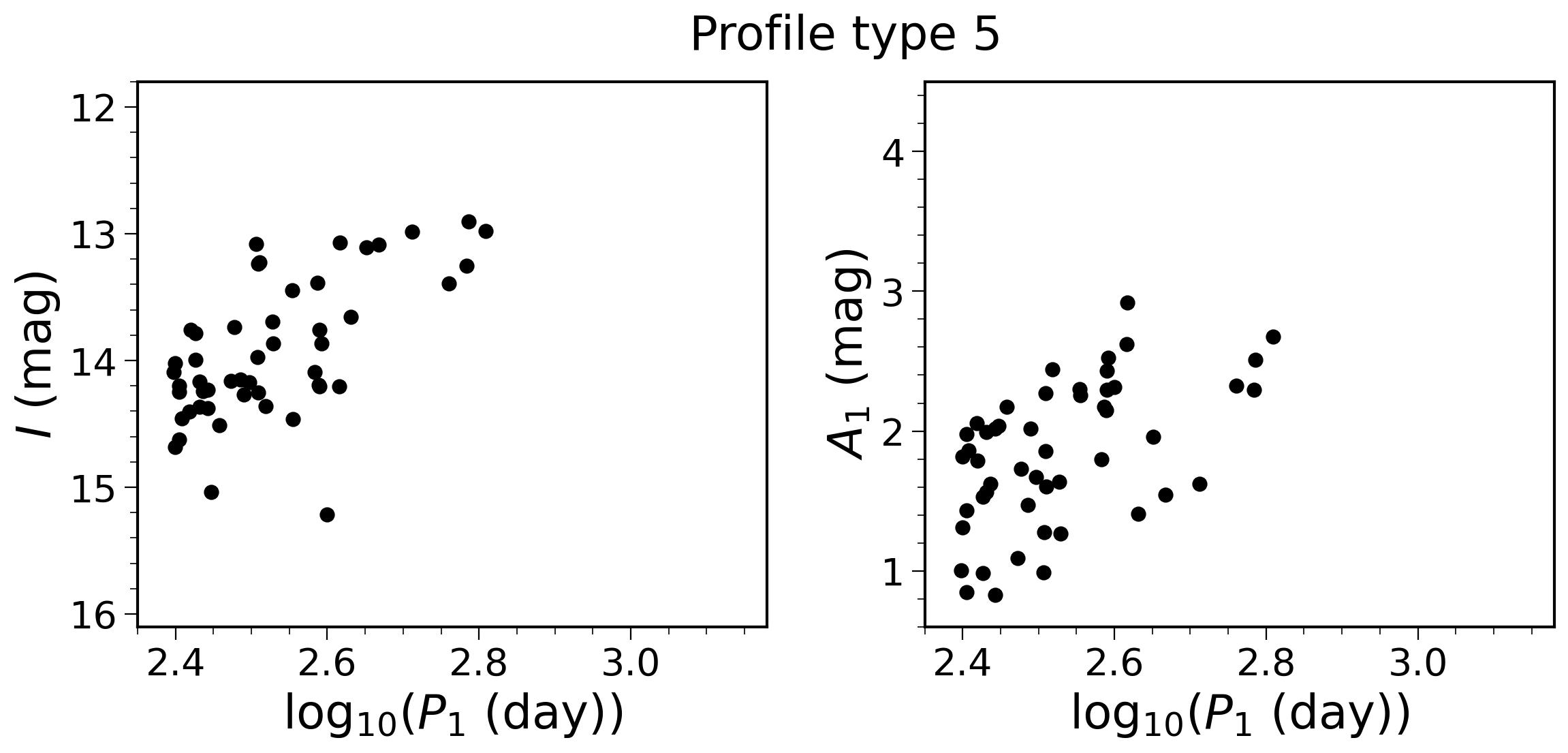}
  \includegraphics[width=0.33\textwidth,trim={0cm 0cm 0cm 0},clip]{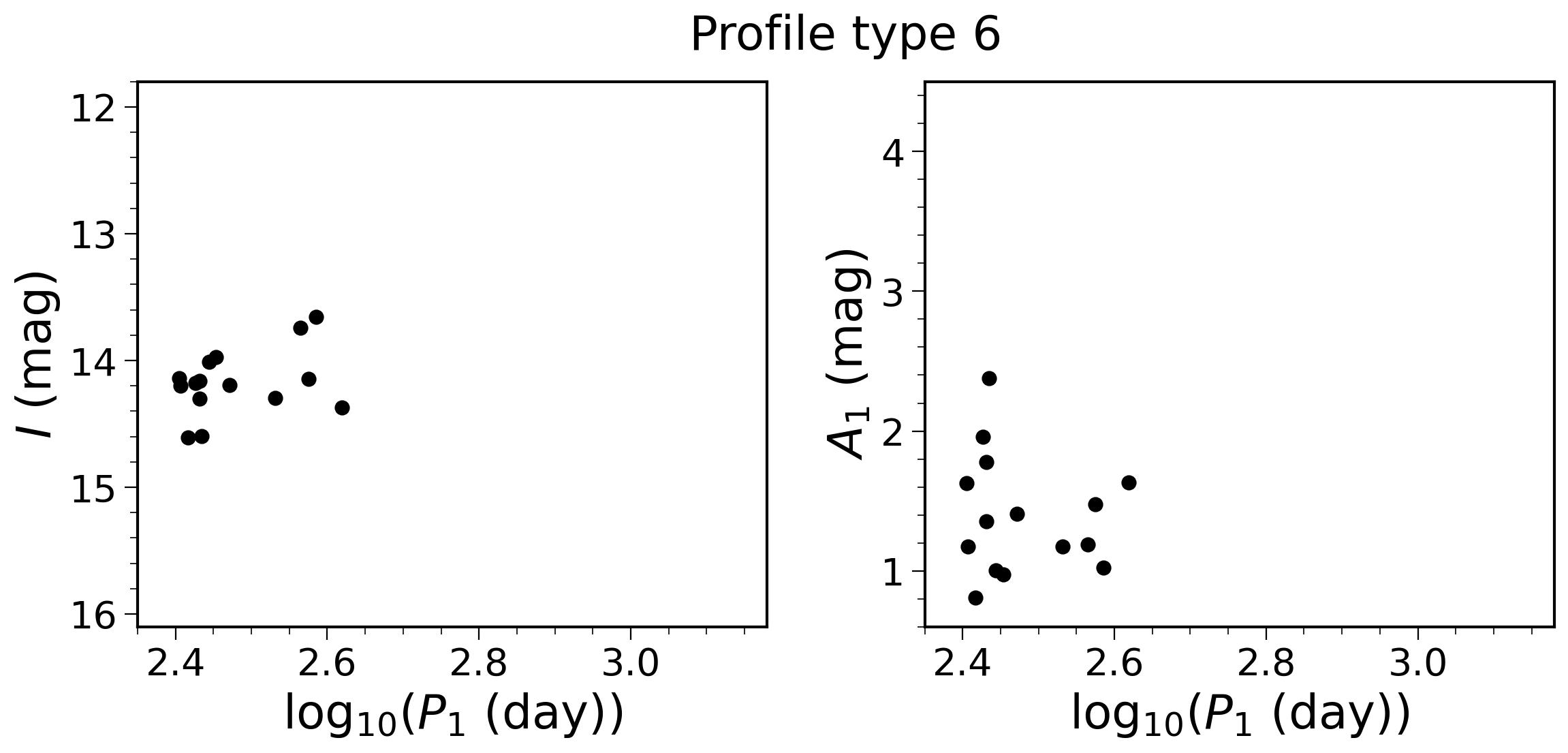}
  \caption{Six pairs of panels, for curves having profile types 1 to 6 as indicated on top of each pair, display the dependence of $I$ and $A_1$ on log$P_1$. The outliers shown in red have $N$$=$89275 (type 2) and $N$$=$66916 (type 4).}
  \label{fig5}
\end{figure*}

As illustrated in Figure \ref{fig5}, different profiles are seen to cover remarkably different domains in the ($I$, log$P_1$) and ($A_1$, log$P_1$) planes, revealing the existence of correlations between the shapes of the curves and the stellar evolution. For each profile type, the curves are confined to well-defined period intervals and to relatively narrow ranges of luminosity and oscillation amplitudes. The only exceptions are curve $N$$=$89275 for profile type 2 and $N$$=$66916 for profile type 4; we ignore them for the time being and describe briefly below the locations covered by each profile type.

Profile type 1 curves cover large periods, in excess of $\sim$400 days and mostly above 650 days; they have large oscillation amplitudes at the mean scale of $\sim$3 mag, slowly increasing with period. As will be later discussed, luminosities are not reliably measured for periods in excess of $\sim$900 days (Figure \ref{fig5}). The mean luminosity is $\sim$14.6 mag for the small group of lower periods and $\sim$13.4 mag for the others.

Profile type 2 curves form a large group, 41 curves, the second in size after profile type 5, 49 curves. Together, they account for more than half of the total sample. They are confined to periods smaller than $\sim$450 days. Between 250 and 450 days, the mean luminosity increases by $\sim$0.7 mag around $\sim$14.4 mag with a dispersion of ±$\sim$0.3 mag. Similarly, the amplitude of oscillation increases from $\sim$2.3 to $\sim$2.8 mag with a same dispersion.

Profile type 3 curves form two small groups separated by a narrow gap; their mean and rms values are $P_1$$=$396$\pm$44 and 632$\pm$34 days, $I$$=$14.3$\pm$0.1 and 13.2$\pm$0.3 mag and $A_1$$=$2.7$\pm$0.3 and 2.8$\pm$0.3 mag, respectively.

Profile type 4 curves are confined to a narrow range of periods, 560$\pm$100 days with large mean luminosities, typically slightly above 13 mag. The amplitude of oscillation increases steeply as a function of period, from less than 1 mag to above 2.4 mag.

Profile type 5 curves form the largest group and behave similarly to profile type 2 curves, covering a same period interval, but with mean luminosities larger by $\sim$0.5 mag and oscillation amplitudes smaller by $\sim$0.8 mag. They include a small group of 4 curves having $P_1$$\sim$630 days, $I$$\sim$13.2 mag and $A_1$$\sim$2.5 mag, which has no equivalent in the profile type 2 group.

Finally, profile type 6 curves are again confined to low periods, below 450 days, with mean luminosities between $\sim$14.6 and $\sim$13.7 mag and oscillation amplitudes below 2 mag.  

\begin{figure*}
  \centering
  \includegraphics[width=0.43\textwidth,trim={0cm -0.05cm 0cm 0},clip]{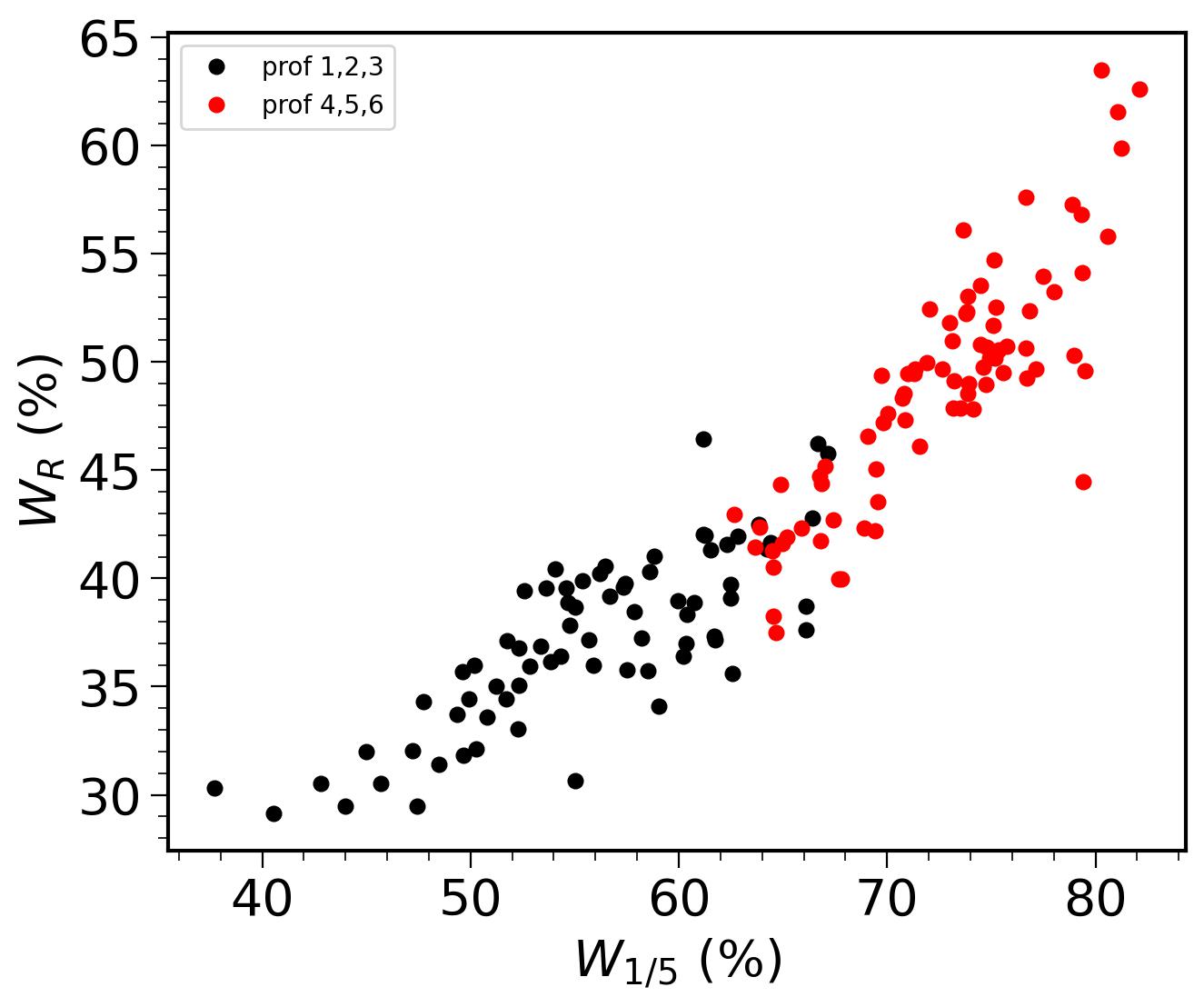}
  \includegraphics[width=0.45\textwidth,trim={0cm 0cm 0cm 0},clip]{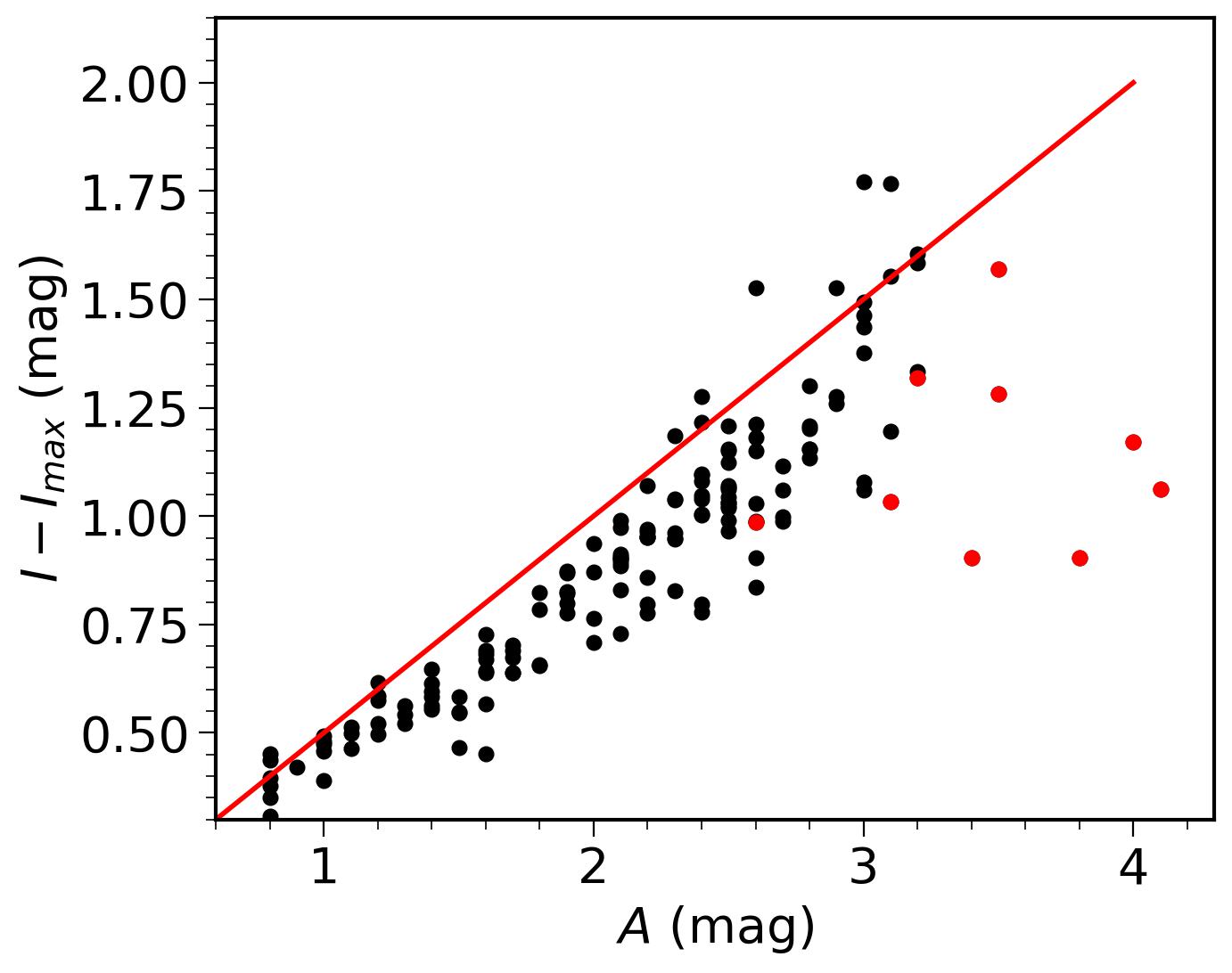}
  \caption{Left: Distribution of the curves of Table \ref{tabA1} in the $W_R$ vs $W_{1/5}$ plane. Different colours indicate different profile types as shown in the insert. Right: dependence of $I-I_{\rm{max}}$ on $A$ for the curves of Table \ref{tabA1}. The curves having periods in excess of 900 days are shown in red. The red line shows the relation $I-I_{\rm{max}}=\frac{1}{2}A$ as reference.}
  \label{fig6}
\end{figure*}

Encouraged by these results, we calculate, for each of the 153 retained curves of Table \ref{tabA1}, the parameters $W_{1/5}$ and $W_R$ that had been introduced in \citet{Nhung2025} to give a measure of the the pulse width. $W_{1/5}$ is the width of a cycle, relative to the period, at an apparent luminosity equal to 20\% of the luminosity amplitude of the oscillation; $W_R$ is the ratio between the luminosity measured above minimum light and integrated over a period and the product of its value at maximum light by the period. The left panel of Figure \ref{fig6} shows that $W_{1/5}$  and $W_R$ are strongly correlated, similarly to what was found in \citet{Nhung2025} for the solar neighbourhood; we therefore  use their geometric mean, $W$$=$$\sqrt{W_{1/5}W_R}$ as a measure of the pulse width from now on, as was done in our earlier work. The values of $W$ are listed in Table \ref{tabA1} together with the phase at minimum light, $\varphi_{\rm{min}}$, which, however, cannot be reliably evaluated when the light maximum is too broad. Moreover, for each curve, Table \ref{tab1} lists values $P$ and $A$ of the period and oscillation amplitude, respectively, obtained from a comparison of the values $P_1$ and $A_1$ listed by OGLE with the values evaluated by visual inspection and those obtained from the evaluation of $W$. In addition to $P$ and $A$, we evaluate $I_{\rm{max}}$ as the I-band luminosity at maximum light, which is slightly more relevant than the mean value $I$ quoted by OGLE \citep{Bhardwaj2019, Bhardwaj2021}. The right panel of Figure \ref{fig6} shows the dependence of $I-I_{\rm{max}}$ on $A$; if one excludes the curves with periods exceeding 900 days, it gives evidence for a good agreement between $I$ and $I_{\rm{max}}$ within uncertainties: their difference is close to half the oscillation amplitude, as expected.  

Being at a much larger distance from us than the solar neighbourhood stars studied in our earlier work, the LMC stars lack the rich set of information offered by the latter. Yet, values of the spectral type and of the visible band luminosity are available and listed in Table \ref{tabA1}. The spectral types are obtained from the SIMBAD data base. Half of the quoted values are M6. For what concerns the V-band luminosity, we use the observations provided by OGLE to evaluate its ratio to the I-band luminosity; for each light curve, we calculate the differences between the $I$- and $V$- band magnitudes measured in a same time bin; their mean and rms values are listed in Table \ref{tabA1} for each light curve separately.

Note that the OGLE data are limited to OGLE-3 observations for the majority of curves but include also earlier OGLE-2 observations for a significant number of them. We treat both sets on an equal footing and we checked that ignoring this difference was not introducing any significant bias and that limiting the study to OGLE-3 observations would not alter any of the conclusions obtained from the present work. 

\subsection{Periods between 250 and 450 days: evidence for two families}
\begin{figure*}
  \centering
  \includegraphics[width=0.9\textwidth,trim={0cm 0cm 0cm 0},clip]{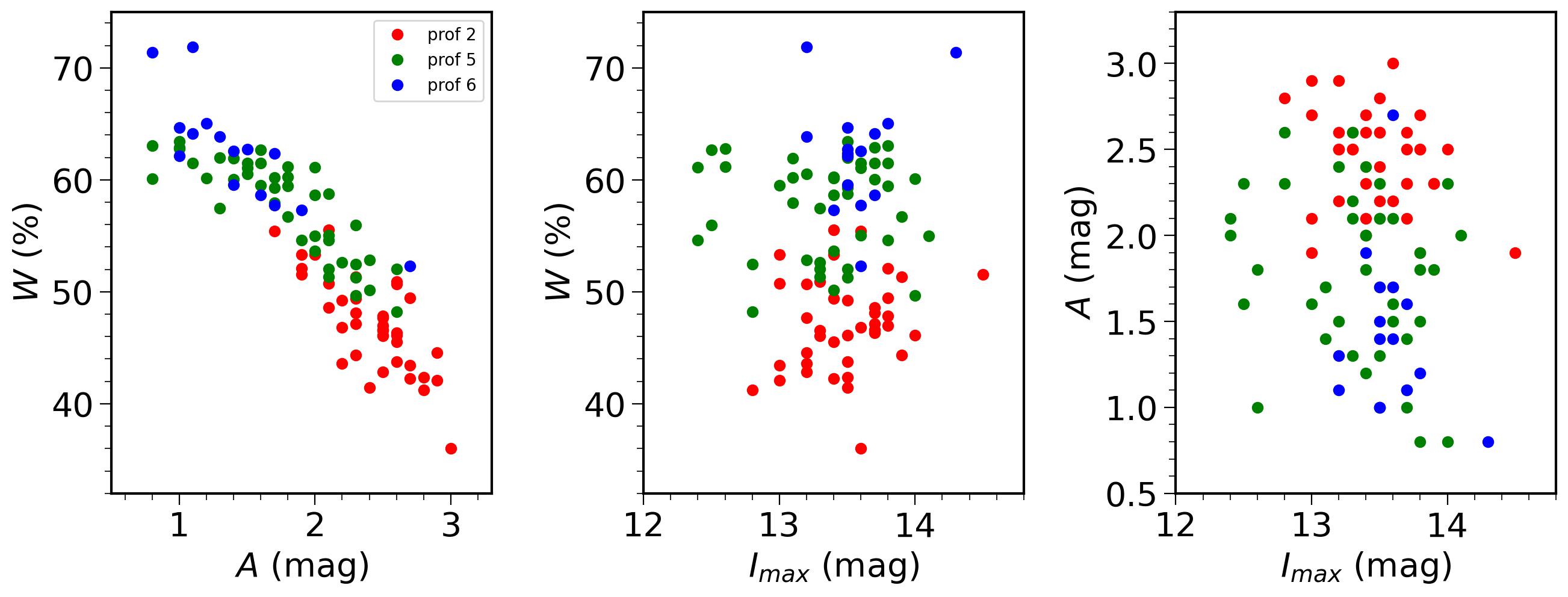}
  \caption{Dependence of $W$ on $A$ (left), $W$ on $I_{\rm{max}}$ (centre) and $A$ on $I_{\rm{max}}$ for the curves having periods between 250 and 450 days. Profile types 2 are coloured in red, profile types 5 in green and profile types 6 in blue.}
  \label{fig7}
\end{figure*}

\begin{figure*}
  \centering
  \includegraphics[width=0.9\textwidth,trim={0cm 0cm 0cm 0},clip]{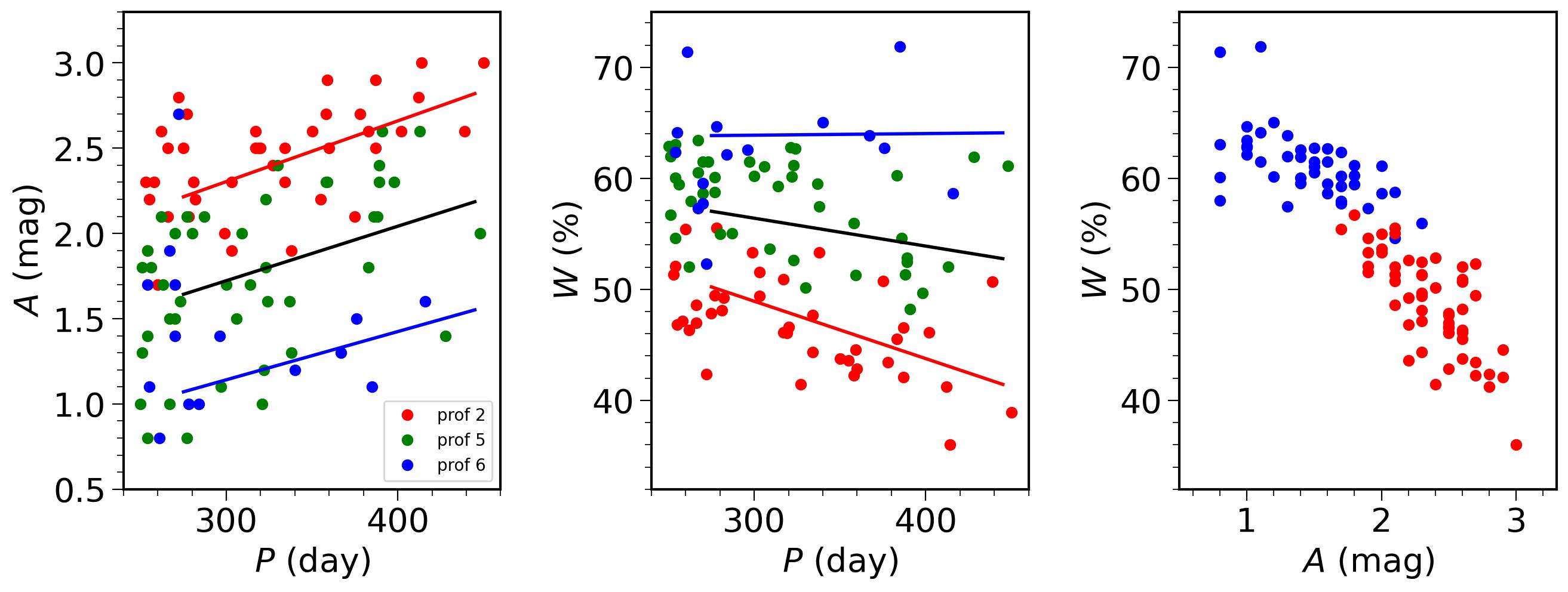}  
  \caption{Dependence on period of $A$ (left) and $W$ (centre) for the curves having periods between 275 and 450 days. Profile types 2 are coloured in red, profile types 5 in green and profile types 6 in blue. The blue and red lines are the results of linear fits to profile types 6 and 2, respectively (Table \ref{tab3}). The black line shows their mean. The right panel show the $W$ vs $A$ distribution of the curves, defining candidates of the A family (red) as having $W$ below the black line in the central panel and candidates of the B family (blue) as having $W$ above the black line in the central panel.}
  \label{fig8}
\end{figure*}

The period interval between 250 and 450 days is nearly exclusively populated by profile types 2, 5 and 6 and corresponds to the region where a difference between the A and B families defined in \citet{Darriulat2026} is most likely to be revealed. We accordingly devote the present sub-section to its study. It includes 115 of the 167 curves of the total sample listed in Table \ref{tabA1}. The dependence of $W$ on $A$, of $W$ on $I$ and of $A$ on $I$ are displayed in Figure \ref{fig7}. Type 6 profiles, with large values of $W$ and small values of $A$, and type 2 profiles, with small values of $W$ and large values of $A$, respectively behave as expected for members of the B and A families defined in \citet{Darriulat2026}. Both cover a same range of $I_{\rm{max}}$, between 13 and 14 mag. Type 5 profiles cover the regions in between and partly overlap each of them. The dependence on period is illustrated in Figure \ref{fig8}. The results of linear fits of the form $A=A_0+A'P$ and $W=W_0+W'P$ show that both display an increase of amplitude and, for type 2 profiles, a decrease of pulse width with increasing period, again as expected for the A and B families. 
  
\begin{table*}
  \centering
  \caption{Parameters of linear fits (275$<$$P$$<$450 days) of the form $A$$=$$A_0$+$A' P$ (mag) and $W$$=$$W_0$+$W'P$ for curves of profile types 2 (upper row) and 6 (lower row).}
  \label{tab3}
  \begin{tabular}{ccccc}
    \hline
&$A_0$& $A'$($\times$10$^3$)&$W_0$&$W'$($\times$10$^2$)\\
Type 2
&1.2$\pm$0.4
&3.6$\pm$1.1
&64$\pm$6
&$-$5$\pm$2 \\
Type 6
&0.3$\pm$0.5
&2.8$\pm$1.4
&63$\pm$10
&1$\pm$3 \\
\hline
\end{tabular}
\end{table*}

These results suggest therefore an assignment of the curves to the A and B families as illustrated in the right panel of Figure \ref{fig8}. While crude and imprecise, such an assignment matches well what is expected and can be adopted with confidence. The mean$\pm$rms values of $V-I$ for 275$<$$P_1$$<$450 days are 4.6$\pm$0.7 mag and 4.0$\pm$0.4 mag for profiles 2 and profiles 6, respectively,  as expected from the larger colour index observed for the Ayes family than for the Byes family in the solar neighbourhood \citep{Darriulat2026}, associated with a larger dust abundance surrounding the star.

Most of the progeny of oxygen-rich stars having periods smaller than 450 days are expected to be carbon-rich, having made the transition at various stages of their evolution, in a range of periods starting as early as 200 days. As mentioned in the next sub-section, only a very few may reach periods larger than 450 days while remaining oxygen-rich Mira variables. But some oxygen-rich stars may never become carbon-rich and spend their whole AGB life as oxygen-rich, as observed for the Blow family in \citet{Darriulat2026}. These are expected to be among the SRV curves of the OGLE data-base illustrated in the upper central panel of Figure \ref{fig2}. 88 of these have periods in excess of 450 days and visual inspection of their light curves shows that all are Long Secondary Periods, which are probably associated with periodic obscuration of carbon-rich stars and will be studied in a forthcoming publication (Nhung et al. in preparation). Probable equivalent of the Blow family defined in \citet{Darriulat2026} are only found with periods typically in the 300 to 450 days range and have mean luminosities in the 12.5 mag region.

\begin{figure*}
  \centering
  \includegraphics[width=0.9\textwidth,trim={0cm 0cm 0cm 0},clip]{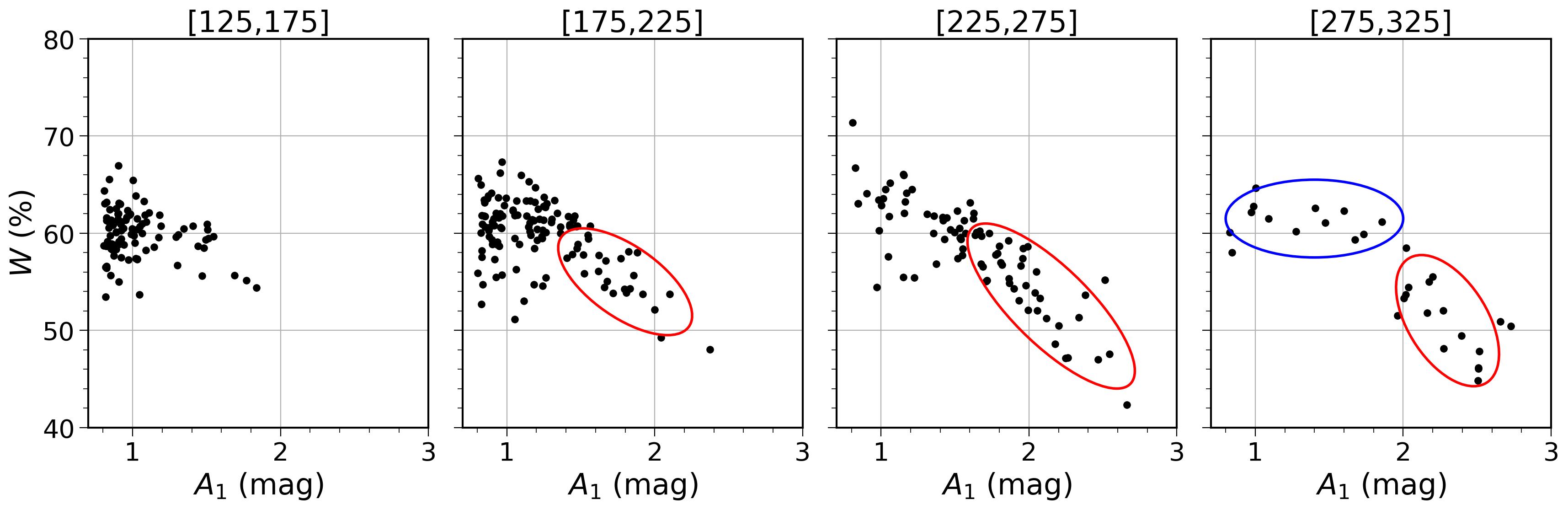}
  \caption{Dependence of $W$ on $A_1$ in different bins of $P_1$ as indicated on top of each panel. The red and blue ellipses illustrate the approximate evolution of the A and B families, respectively. }
  \label{fig9}
\end{figure*}

The progenitors of the Mira oxygen-rich variables studied in the present sub-section are expected to have periods smaller than 250 days and values of $W$, $A$ and $I_{\rm{max}}$ in continuity with theirs. Those assigned to the A family are expected to maintain large, but decreasing, oscillation amplitudes as their period decreases below 200 days and their pulse width is expected to increase. Those assigned to the B family are instead expected to experience a faster decrease of the oscillation amplitude and a more modest variation of the pulse width. This is indeed consistent with what is observed: Figure \ref{fig9} displays the dependence of $W$ on $A_1$ in different bins of $P_1$ for the curves having periods between 100 and 250 days; it shows clearly that, as the period decreases, the curves of the A family drift toward lower values of $A_1$ and larger values of $W$, as expected for family A while those of the B family keep $W$ approximately constant while their amplitude decreases promptly below $\sim$1 mag, as expected for family B. For periods smaller than $\sim$200 days, the former can still be identified as clear pulsators with close to sine wave light curves of amplitudes smaller than 2 mag, while the latter have become irregular with amplitudes smaller than 1 mag. 

\subsection{Periods larger than 450 days}

Over three decades ago, \citet{Wood1992} and \citet{Zijlstra1995} reported observations of OH/IR stars in the LMC. These are evolved oxygen-rich AGB stars which exhibit OH maser emission at 1612 MHz and are significantly obscured by dust; this makes them opaque at visible wavelengths and bright in the infrared, the stellar light being converted to larger wavelengths by the dust grains. They are at the end of their AGB life and experience large mass loss rates, typically larger than 10$^{-6}$ M$_{\odot}$\,yr$^{-1}$. A decade later, \citet{Whitelock2003} reviewed observations of obscured stars in the LMC and evaluated their bolometric luminosities $M_{\rm{bol}}$ using observations in the JHKL bands from the 1.9 m SAAO telescope, IRAS and ISO. In particular, they confirmed the presence of a group of stars with periods larger than 1000 days and $M_{\rm{bol}}$$\sim$$-$7 mag and of HBB stars with periods between 500 and 700 days and $M_{\rm{bol}}$$\sim$$-$6 mag. Recently, \citet{Olofsson2022} and \citet{JimenezEsteban2021} have reported observations of OH/IR stars in the Galaxy. \citet{Goldman2017} detected 12 OH maser sources out of a sample of 17 OH/IR LMC candidates with $<$$P$$>$$=$1131 days, $<M_{\rm{bol}}>$$=$$-$7.13 mag and an average spectral type M7.6 (their Table 2) and studied the dependence on metallicity of the wind speeds, dust content, and mass-loss rates.

It is therefore compelling to interpret the 9 Table \ref{tabA1} stars having periods larger than $\sim$900 days (1103 days on average) as possible OH/IR candidates. They all have profile type 1, with a large value of $\varphi_{\rm{min}}$ (0.65 on average) revealing the steep ascending branch, and large oscillation amplitudes (3.5 mag on average). Only three of them could be observed in the visible, with very large values of $<$$V-I$$>$, 7.3, 7.7 and 4.7 mag, respectively; the six others are probably too strongly obscured to be detected in the visible. Five have spectral types quoted in the SIMBAD data base, all larger than M7, M8.7 on average. All these features match well what is expected from OH/IR stars and support such an assignment. Periods between 700 and 900 days host a group of 5 other curves of profile type 1, in clear continuity with the OH/IR curves, which makes them good candidates for being their progenitors. Indeed, \citet{Whitelock2003} remarked that, perhaps surprisingly, most of the LMC very long-period obscured stars seem to follow an extrapolation of the period–luminosity relation determined for stars with shorter periods, consistent with what we observe.

As was said above, the period interval below 700 days is host to HBB stars. \citet{Ventura2015} and \citet{GarciaHernandez2009} have presented arguments pleading for an important part of the population of obscured LMC stars to be of this type. The 14 curves of profile type 4 having a period larger than 460 days display a double maximum reminiscent of that displayed by the Galaxy HBB archetypes R Cen and R Nor \citep{Nhung2025} and are obvious candidates for such an assignment. They have large luminosities at maximum light, 12.2 mag on average, and a mean period of 555 days. They are likely candidate progenitors of the 12 curves having periods between 570 and 690 days, 625 days on average, 8 of profile type 3 and 4 of profile type 5. These also have large luminosities at maximum light, 12.0 mag on average, and have reached large oscillation amplitudes, 2.7 mag on average. All have profiles close to double maxima, supporting the HBB assignment.

This leaves only 7 stars having periods larger than 450 days and not being HBB or OH/IR candidates. Their mean period is only 498 days, suggesting that they are the only progeny of the 115 stars of the A and B families identified in the preceding sub-section. This implies that the vast majority of the stars of the A and B families have become carbon-rich before reaching a period of 450 days. We defer their study to a later publication (Nhung et al., in preparation).

\begin{figure*}
  \centering
  \includegraphics[width=0.7\textwidth,trim={0cm 0cm 0cm 0},clip]{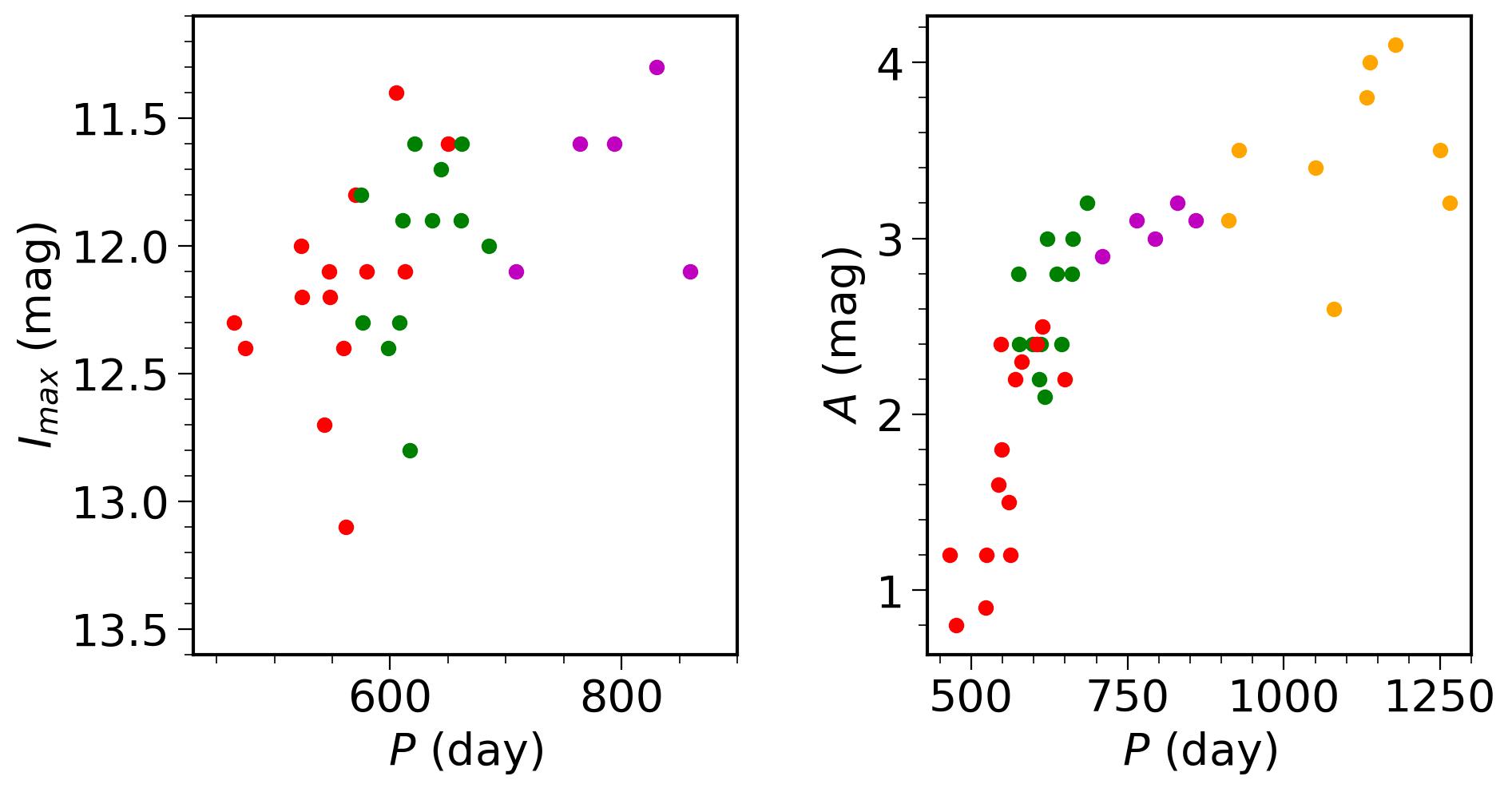}
  \caption{Distribution of the large period curves in the $I_{\rm{max}}$ vs $P$ (left) and $A$ vs $P$ (right) planes. The 9 OH/IR candidates, which have profile 1, are shown in orange in the right panel. Their likely progenitors (5 curves having also profile 1), are shown in purple. The 14 HBB candidates, which have profile 4, are shown in red in both panels and their likely progeny (12 curves of profiles 3 and 5) are shown in in green.}
  \label{fig10}
\end{figure*}

In summary, the properties of the light curves of LMC oxygen-rich Mira variables having a period larger than 450 days suggest that 14 of these stars are OH/IR stars and 26 are HBB stars, leaving only 7 stars as likely progeny of the lower period stars of the A and B families. Their distributions in the $I_{\rm{max}}$ vs log$P$ and $A$ vs log$P$ planes are illustrated in Figure \ref{fig10}. 

\section{Summary and conclusion}
We presented a study of the shapes of the OGLE light curves of oxygen-rich Mira variables in the LMC \citep{Soszynski2009a} with the aim to compare them with such curves observed in the solar neighbourhood. An earlier study of the latter \citep{Darriulat2026} had revealed the presence of two different families of curves associated with two different paths of evolution on the AGB: the question was therefore whether two analogous families could be identified in the LMC. We started by exploring a region of the period vs amplitude plane where the distinction between the two families had first been revealed in the solar neighbourhood: Mira variables having periods between 250 and 450 days. Visual inspection of the light curves listed by OGLE in this region proved to be an essential preliminary to the study. It showed that the curves, in their vast majority, display clear and relatively regular pulsations allowing for a reliable evaluation of their period and amplitude, which the OGLE parameters $P_1$ and $A_1$ were found to properly describe. Moreover, it gave evidence for the presence of three different shapes of the pulsation cycle, referred to as profiles 2, 5 and 6, populating different regions of the period vs amplitude plane, which turned out to be a very efficient tool to identify the equivalent of the A and B families in the LMC.

In addition to the detailed study of this sample of Mira variables of moderate periods, we also explored regions of smaller and larger periods. The former made it possible to identify progenitors of the main sample, the latter to identify both HBB and OH/IR candidates.

The OGLE observations, in the I-band, that we use for the LMC are of a very different nature from the AAVSO observations, in the visible, that were used for the solar neighbourhood. Their main assets are a high density of observations, typically one every fifth day, with excellent sensitivity and accuracy of the luminosity measurement, typically less than a hundredth of magnitude; and a unique and well-known distance from Earth, implying that the measured luminosity provides directly a measure of the absolute luminosity. Their main drawback is the presence of significant gaps in the sequence of observations. The limitation of the total observing time to a decade or so, an order of magnitude smaller than for the AAVSO observations, is a serious drawback for the study of low amplitude semi-regular variables but not for the Mira variables studied here, for which several cycles are observed. Comparisons of the OGLE photometry with that available at other infrared wavelengths in the published literature has made it possible to study the colour properties of the stars and has been extensively made use of, in particular for ascribing each curve to the O-rich or C-rich group. But detailed observations available in the solar neighbourhood, both in the radio range with instruments such as ALMA and in the visible/near infrared range with instruments such as the VLT are of course lacking for the LMC, implying the absence of measurement of quantities such as the stellar radius and the mass-loss rate.

These different specificities of the OGLE and AAVSO observations have a significant impact on the results of the present study. While there is no doubt that we have been able to identify with confidence in the LMC two different families of curves that display properties similar to those that had been previously identified in the solar neighbourhood, the image that could be drawn is somewhat rough and approximate. No argument was found that could contradict our conclusions but the arguments that we were able to produce in order to plead in their favour are not sufficiently strong to reveal by themselves, with enough confidence, the existence in the LMC of two clearly distinct families as has been done in the solar neighbourhood. The presence in the LMC of two families having similar properties to those of the solar neighbourhood is an important result that gives increased confidence in the conclusions presented in \citet{Darriulat2026} for the solar neighbourhood; but it asks for new efforts aimed at establishing their validity with increased robustness. A first step in this direction is to extend the study to the samples of carbon-rich and semi-regular curves, which we have undertaken and the results of which will be presented in a future publication.

\begin{acknowledgements}
This research has made use of the SIMBAD database, CDS, Strasbourg Astronomical Observatory, France. It is funded by the Vietnam National Foundation for Science and Technology Development (NAFOSTED) under grant number 103.99-2024.36.  Financial support from the World Laboratory, the Rencontres du Vietnam and the Vietnam National Space Center is gratefully acknowledged. 
\end{acknowledgements}

\bibliography{lightcurves5}{}
\bibliographystyle{raa}

\appendix

\section{}

\begin{table*}[h]
  \centering
  \caption{List of the 167 curves selected from the OGLE data base as LMC/Mira/AGB/oxygen-rich, $P_1$$>$250 days. The first column gives the ranking, $n$, in the list. The second column gives the number $N$ in the OGLE identification $n$.OGLE-LMC-LPV-$N$. The third, fourth and fifth columns give the I-band luminosity at maximum light, $I_{\rm{max}}$, the period $P$ (in days) and the amplitude of oscillation, $A$ (in units of magnitude), as evaluated by us (Section 2.2). The sixth column gives the shape of the pulse profile as illustrated in Figures \ref{fig4} and \ref{figA1} to \ref{figA7} below. Rejected curves are given a 0 shape. The seventh and eight columns give the phase at minimum light, $\varphi_{\rm{min}}$, and the pulse width, $W$, as evaluated by us (Section 2.2); in a few cases, the pulse is too broad to allow for a reliable evaluation of $\varphi_{\rm{min}}$. The ninth column gives the mean and rms values of the difference between visible and I-band magnitudes as measured by OGLE when this evaluation can be made reliably. The tenth column gives the spectral type when available from the SIMBAD data base. The full table is available at https://www.scidb.cn/en/s/UvMzeu.}
  \label{tabA1}
  \begin{tabular}{ccccc ccccc}
    \hline
    $n$ &$N$& $P$&$I_{\rm{max}}$	&$A$&	Shape&	$\varphi_{\rm{min}}$	&$W$	&$V-I$&	Sp. \\
    \hline
    1&	00902&	296&	13.6&	1.4&	6&	0.44&	63&	3.8/0.9&	$-$	\\
2&	00990&	321&	12.6&	1.0&	5&	0.32&	63&	2.7/0.6&	$-$	\\
3&	01312&	1179&	13.8&	4.1&	1&	0.64&	44&	$-$&	M7.5	\\
4&	01998&	256&	$-$&	$-$&	0&	$-$&	$-$&	$-$&	$-$	\\
5&	03910&	1250&	12.0&	3.5&	1&	0.56&	40&	7.7/0.9&	$-$	\\
6&	04312&	1265&	12.9&	3.2&	1&	0.76&	39&	7.3/0.7&	M10	\\
7&	04669&	416&	13.7&	1.6&	6&	$-$&	59&	4.4/0.5&	$-$	\\
8&	04689&	1080&	13.7&	2.6&	1&	0.8&	47&	$-$&	M9	\\
9&	04798&	360&	13.2&	2.5&	2&	0.52&	43&	6.3/0.7&	$-$	\\
10&	05179&	756&	$-$&	$-$&	0&	$-$&	$-$&	$-$&	$-$	\\
...\\
\hline
\end{tabular}
\end{table*}

\begin{figure*}[h]
  \centering
  \includegraphics[width=1\textwidth,trim={0cm 0cm 0cm 0},clip]{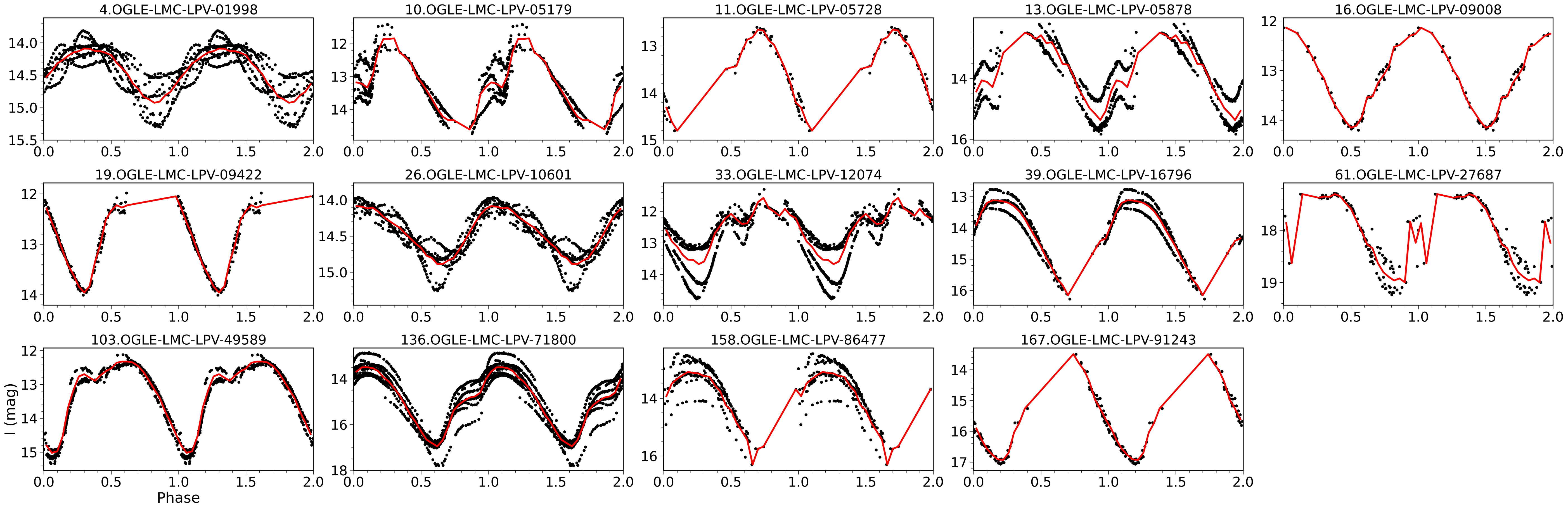}
  \caption{Shape profiles of the 14 curves listed in Table \ref{tabA1} that have been rejected for having data of insufficient quality. The mean profile is shown in red.}
  \label{figA1}
\end{figure*} 

\begin{figure*}
  \centering
  \includegraphics[width=1\textwidth,trim={0cm 0cm 0cm 0},clip]{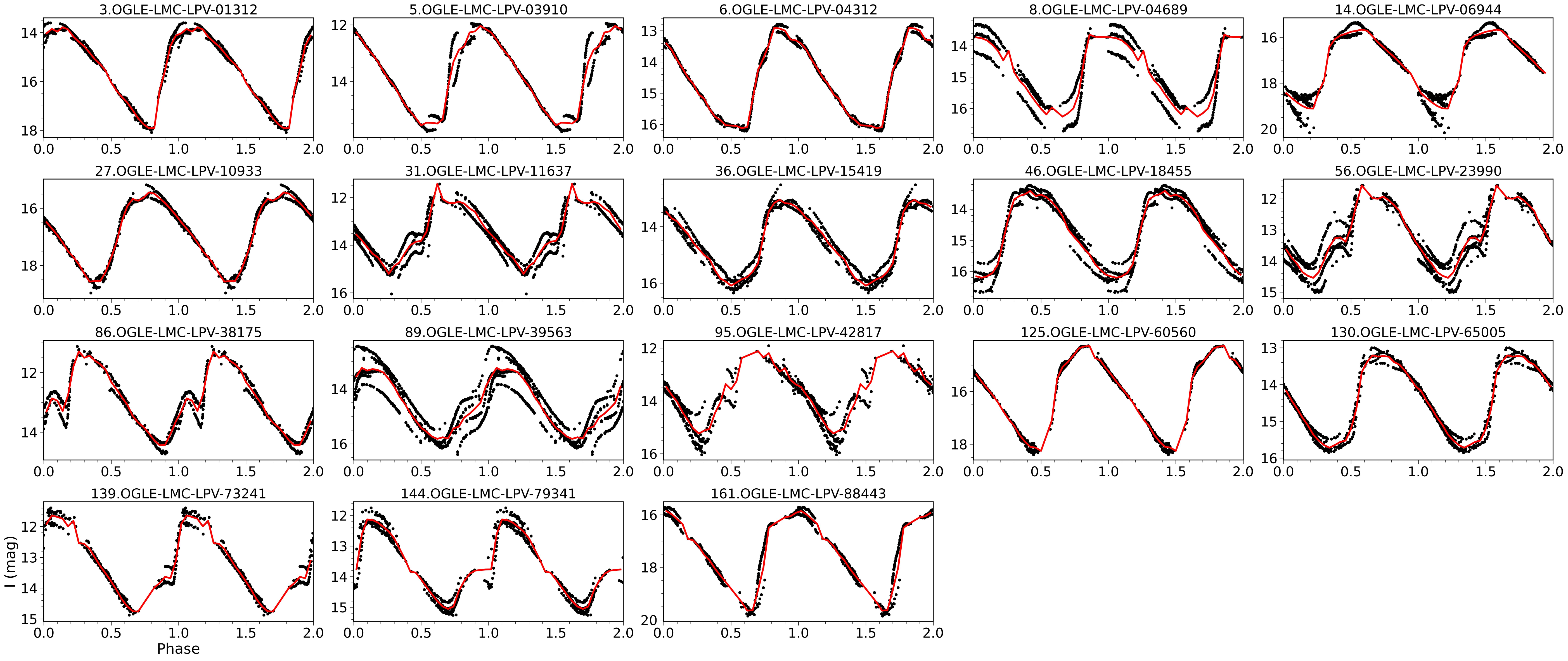}
  \caption{Same as Figure \ref{figA1} for the 18 profiles of type 1.}
  \label{figA2}
\end{figure*} 

\begin{figure*}
  \centering
  \includegraphics[width=1\textwidth,trim={0cm 0cm 0cm 0},clip]{figA3.jpg}
  \caption{Same as Figure \ref{figA1} for the 41 profiles of type 2.}
  \label{figA3}
\end{figure*} 

\begin{figure*}
  \centering
  \includegraphics[width=1\textwidth,trim={0cm 0cm 0cm 0},clip]{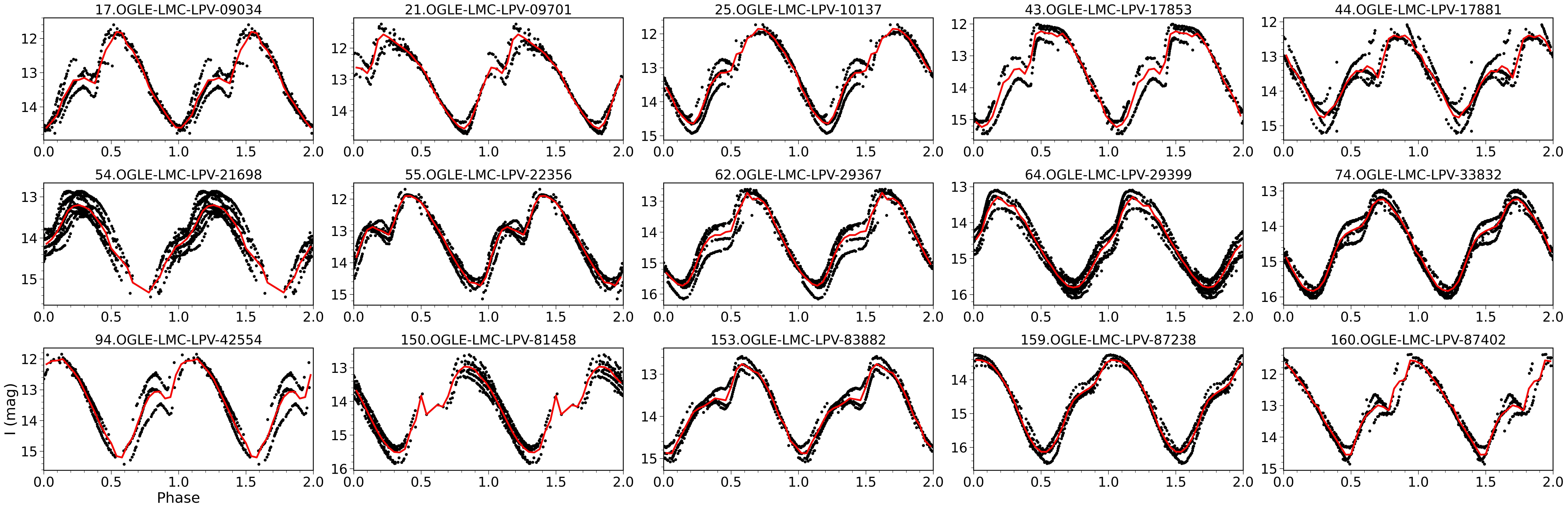}
  \caption{Same as Figure \ref{figA1} for the 15 profiles of type 3.}
  \label{figA4}
\end{figure*} 

\begin{figure*}
  \centering
  \includegraphics[width=1\textwidth,trim={0cm 0cm 0cm 0},clip]{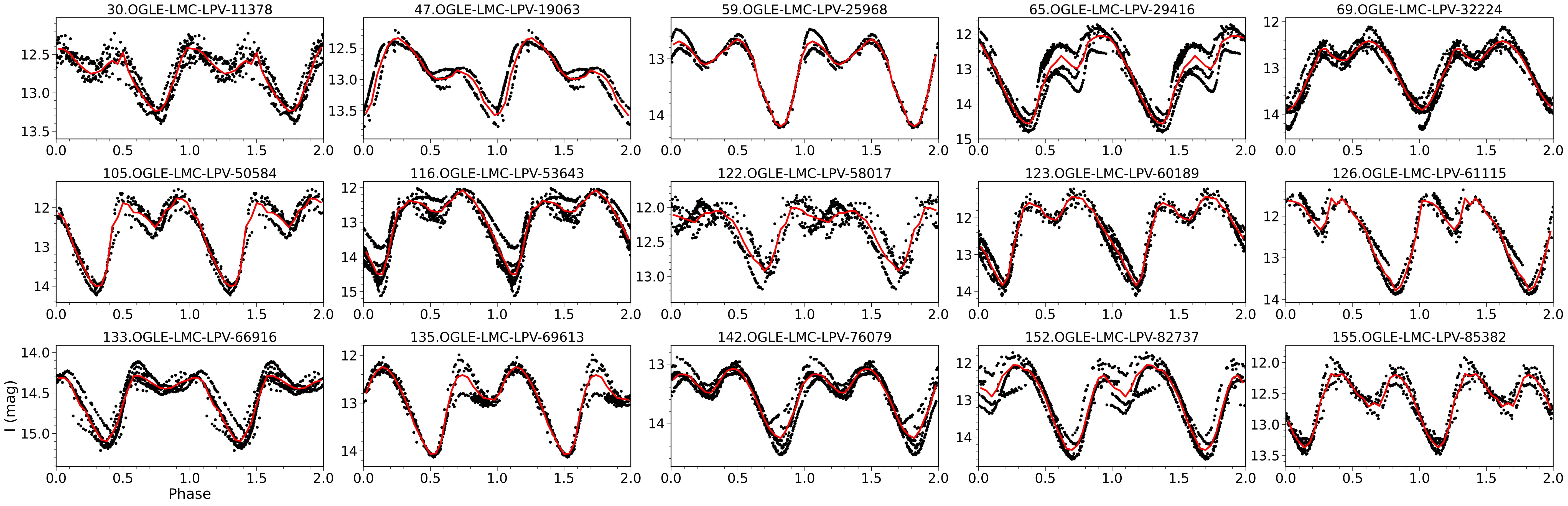}
  \caption{Same as Figure \ref{figA1} for the 15 profiles of type 4.}
  \label{figA5}
\end{figure*} 

\begin{figure*}
  \centering
  \includegraphics[width=1\textwidth,trim={0cm 0cm 0cm 0},clip]{figA6.jpg}
  \caption{Same as Figure \ref{figA1} for the 49 profiles of type 5.}
  \label{figA6}
\end{figure*}

\begin{figure*}
  \centering
  \includegraphics[width=1\textwidth,trim={0cm 0cm 0cm 0},clip]{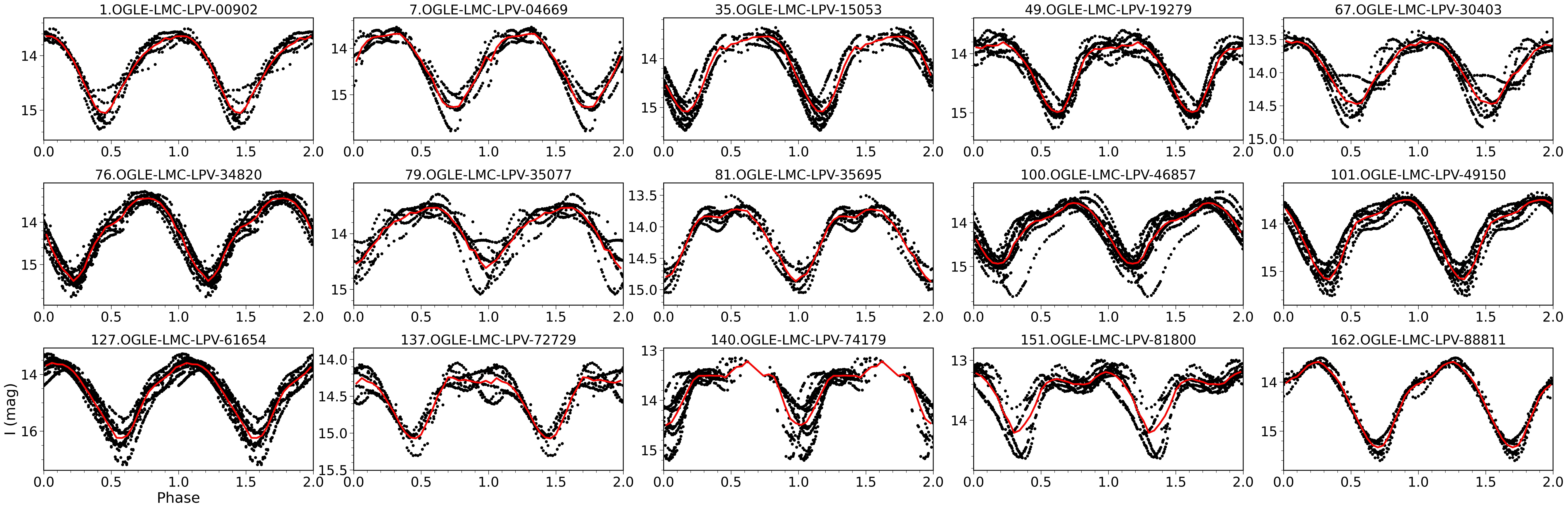}
  \caption{Same as Figure \ref{figA1} for the 15 profiles of type 6.}
  \label{figA7}
\end{figure*} 

\label{lastpage}

\end{document}